%% file: AniCode.tex
\documentclass[acmtog]{acmart}
\renewcommand\footnotetextcopyrightpermission[1]{} 
\usepackage{amsmath}
\usepackage{bm}
\usepackage{booktabs} 

\setcitestyle{square}

\usepackage[ruled]{algorithm2e} 

\SetAlFnt{\small}
\SetAlCapFnt{\small}
\SetAlCapNameFnt{\small}
\SetAlCapHSkip{0pt}
\IncMargin{-\parindent}

\begin{document}
\title{AniCode: Authoring Coded Artifacts for Network-Free Personalized Animations}

\author{Zeyu Wang, Shiyu Qiu, Qingyang Chen, Alexander Ringlein,\\Julie Dorsey, Holly Rushmeier}
\affiliation{
 \institution{Yale University}
 \streetaddress{51 Prospect St}
 \city{New Haven}
 \state{CT}
 \postcode{06511}
 \country{USA}}
\authorsaddresses{}
\renewcommand\shortauthors{Wang, Z. et al}

\begin{abstract}
Time-based media (videos, synthetic animations, and virtual reality experiences) are used for communication, in applications such as manufacturers explaining the operation of a new appliance to consumers and scientists illustrating the basis of a new conclusion. However, authoring time-based media that are effective and personalized for the viewer remains a challenge. We introduce AniCode, a novel framework for {\em authoring} and {\em consuming} time-based media.  An author encodes a video animation in a printed code, and affixes the code to an object. A consumer uses a mobile application to capture an image of the object and code, and to generate a video presentation on the fly. Importantly, AniCode presents the video personalized in the consumer's visual context. Our system is designed to be low cost and easy to use. By not requiring an internet connection, and through animations that decode correctly only in the intended context, AniCode enhances privacy of communication using time-based media. Animation schemes in the system include a series of 2D and 3D geometric transformations, color transformation, and annotation. We demonstrate the AniCode framework with sample applications from a wide range of domains, including product ``how to'' examples, cultural heritage, education, creative art, and design. We evaluate the ease of use and effectiveness of our system with a user study.

\end{abstract}

%
%

%
%


\begin{teaserfigure}
  \includegraphics[width=\textwidth]{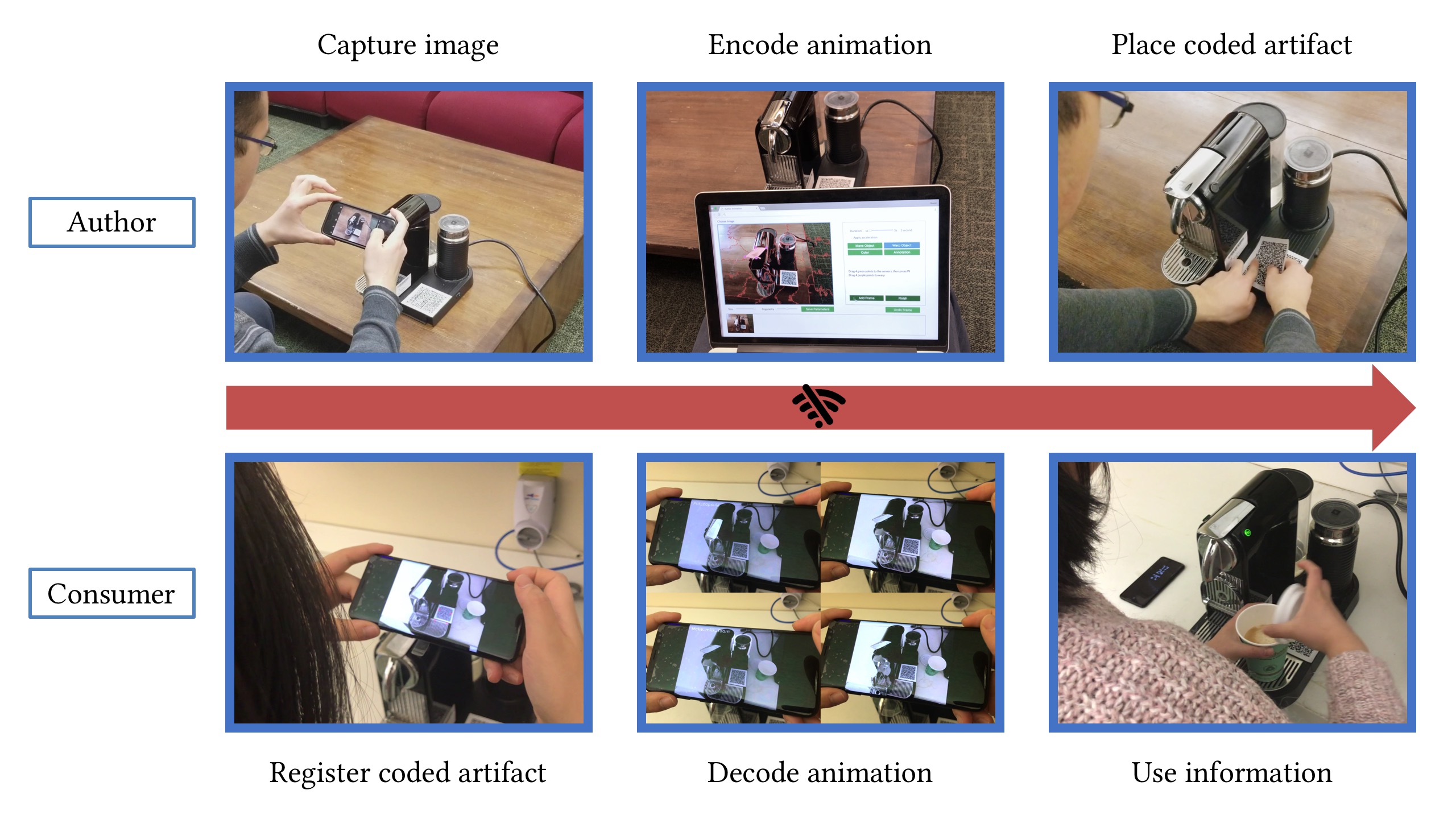}
  \caption{Authoring and generating network-free personalized animations. The author takes a picture, specifies parameters and regions to be animated, and encodes the information in a QR code following designed animation schemes. With the authored QR code, our mobile application on the consumer's side registers the view and decodes a personalized animation. The consumer obtains visual information such as how to use a coffee maker without having to access the internet.}
  \label{fig:teaser}
\end{teaserfigure}

\maketitle
\thispagestyle{empty}

\input{1_introduction}

\input{2_related_work}

\input{3_framework_overview}

\input{4_methodology}

\input{5_sample_applications}

\input{6_evaluation}

\input{7_limitations_and_future_work}

\bibliographystyle{ACM-Reference-Format}
\bibliography{bibliography}

\end{document}

%% file: 1_introduction.tex
\section{Introduction}
\label{sec:1}

Time-based media, such as videos, synthetic animations, and virtual reality experiences increasingly are replacing text and static images in communication.
The popularity of this particular type of communication is evident from the ubiquity of 
do-it-yourself ``how to'' videos, technical lectures on public platforms, and virtual tours presented as panoramic videos. Advances in software and hardware make it possible to render high-quality animations and capture high-resolution digital videos.
However, 
production of useful content requires time and expertise to shoot and edit video clips, or to model synthetic objects for rendering. 
Furthermore, the same content is created for all consumers, and lacks personalization. Current video-based communication also requires network accessibility and abundance of storage. While widespread availability of high-speed internet makes it easy to watch large videos almost anywhere with internet access, there are no alternatives for video-based communication in environments with little or no network access. Furthermore, in an age where personal data is easily collected and tracked, there is increased need for applications that protect privacy and do not rely on network access.

We propose the AniCode framework as an alternative to traditional video generation and sharing. We distinguish between {\em authors}, who have content to be communicated, and {\em consumers} who view and use the content (Fig.~\ref{fig:teaser}). Our system enables authors to {\em create} coded artifacts that allow consumers to {\em generate} time-based displays using their own device. One example is authoring a customized demonstration of the operation of a complicated piece of equipment that is already installed at a consumer's site. An animation is far more effective than the printed or electronic manual with text and a series of still images.
In this scenario, a simple smartphone camera can capture a code embedded in the equipment. We use the code to match image segments and apply authored transformations to generate the animation that shows how to operate the equipment. Since the imagery is in the consumer's own visual context, the motion of objects is clear to the consumer. Due to matching of segments, the correct animation can only be generated when the consumer is in the intended context, thus helping to enhance privacy. Communication is also more private and robust to network instability by generating the animation on the fly without any need to access a network. To focus on preserving the key components of the animation, as well as leaving room for the viewer to have a more personalized experience, a limited amount of information is encoded. Our framework makes use of a consumer's own photograph, and focuses on communicating ideas rather than polished visuals. This allows the production of a video typically 20 MB in size from 200 bytes of coded animation -- effectively a 100,000 compression ratio.


Our work makes the following contributions:
\begin{itemize}
\item 
We propose an alternative framework for video-based communication based on authoring a coded artifact that is imaged to create an animation on demand.
\item We show that the new framework offers enhanced (but not absolute) robustness and privacy by not relying on the presence of an internet connection and requiring consumers to be on-site and in the correct context to be able to view the animation. 
\item We present an authoring interface that enables non-experts to author animations and create augmented reality content. We conduct a user study to evaluate the usability of the interface.
\item We present 
a consumer-side 
mobile application that allows the consumer to image a coded-artifact 
and reconstruct the authored animation 
on the fly in their own visual context.
\item We show how this framework improves the documentation of products, educational sequences, and offers a new methodology for presentation of scientific results to the general public 
with sample applications drawn from several fields.
\end{itemize}

%% file: 2_related_work.tex
\section{Related Work}
\label{sec:2}

At a high level, our work builds on a long history of generating animations in computer graphics \cite{parent2012computer}, authoring multimedia \cite{feiner1991automating}, and computer graphics applied to creating expository media \cite{Agrawala:2003:DES:882262.882352,Agrawala:2011:DPV:1924421.1924439}. We rely on recent work on communicating using time-based media, on time-based media authoring systems, on mobile communication methods assisted by augmented reality, and on network-free communication systems.

\subsection{Time-based media} 
Time-based media, such as video, can be more effective than text-based media in education \cite{Schnotz2002}, entertainment \cite{karat2001less}, and advertising \cite{appiah2006rich}. People believe that animations optimally employ the audience's cognitive capacity \cite{wouters2008optimize} and promote understanding of concepts \cite{barak2011learning}. Producers \cite{clarine} and analysts of social media \cite{kaplan2010users} describe video as powerful because it can incorporate other media, present rich content, motivate action, enhance social communication, enable convenient and accessible production, and reach a wide audience. Digital technologies for creating time-based media are now broadly accessible and frequently applied to tell stories. In cultural heritage, for example, it is well-recognized that any data needs to be presented with context, such as historical notes, physical descriptions, stylistic analysis, and other information \cite{murcher}. Storytelling from data is an important motivation for the system described in this paper, but our work is not on the humanist or psychological principles of what makes a story effective, but rather on creating a new framework for authors to apply those principles.
 
\subsection{Authoring Systems}
Authoring for multimedia systems has progressed since seminal work such as the COMET system \cite{feiner1991automating}. One recent line of research investigates the possibilities for improving the authoring and presentation of video tutorials. This work includes a study of authors of ``how to'' videos and a development of an improved system for authoring. This system includes the ability to add bookmarks to a captured video, and then add multimedia annotations to the video \cite{Carter2014}. Other work also considers the process of effective video capture. For example, an improved first-person viewpoint is developed using Google glass for capture \cite{carter2015creating}. On a high level, this work emphasizes the importance of annotation systems, linked diverse data types, and reconsidering content capture.

Animation generation has long been a feature of information and scientific visualization systems \cite{31462}. Recent work in scientific visualization considers the authoring of 360-degree videos with branching \cite{Liao2014,chu2017navigable}. This work emphasizes both the design of the authoring and navigation interfaces.
There are also research projects that focus on authoring tools for end users in mobile applications \cite{7388029}, which keeps more people in the loop of content production.

\subsection{Augmented Reality}
The popularization of virtual reality (VR) and augmented reality (AR) provides a new approach to communication. Content creation is one of the key areas of AR development. Some applications focus on superimposing virtual objects on a real environment based on coplanarity in single images \cite{cho2016content}. Similar work includes manipulating existing objects interactively with partial scene reconstructions based on cuboid proxies \cite{zheng2012interactive}. Authoring time-based media for AR has also become possible by mixing raw video footage using sparse structure points \cite{chang2016interactive}. Recent AR devices such as Microsoft HoloLens provide the possibility of interactive scene editing in the real world based on data from the sensor \cite{yue2017scenectrl}. Some industrial products can create AR tutorials from normal videos, offering users a personalized experience \cite{IOXP}. In particular, smartphones with high-resolution cameras provide extraordinary resources for acquiring and processing visual information, thus making mobile VR and AR possible. Researchers also exploit mobile access to context-based information in many sites that require interactive expository resources, where smartphones are used to receive input from the user and create augmented environments \cite{andolina2012exploitation}.

These systems have interesting insights into content creation for AR. However, they require the author to explicitly transfer a large amount of data to the consumer via internet or local copy in order to communicate ideas. In this paper, we develop methods for users to access personalized animations without an internet connection, which makes the communication more robust to network instability and helps preserve the user's privacy. Compared with prior AR systems, the AniCode framework augments user-captured images to produce animations locally and offers a valuable tool to provide information and engage users without relying on a communication network. Our mobile application is not confined to a specific scene. 

\subsection{Network-Free Communication}
Data privacy has become an important focus in systems and applications. When a smartphone is connected to a public network, data privacy is at risk. For instance, users of the fitness tracking app Strava unwittingly gave away locations of secret U.S. army bases \cite{Strava}. We explore communicating information visually without internet connection. Our work can be considered a variation of ``Visual MIMO'' \cite{ashok2014}. Optical communication systems are designed using active light emitting systems and cameras. 
Methods for manipulating light emitting systems and then using computer vision techniques on the receiver end have been explored extensively \cite{ashok2011}. Additional techniques such as AirCode \cite{li2017aircode} and FontCode \cite{xiao2018fontcode} embed information in a network-free environment. AirCode exploits a group of air pockets during fabrication and the scattering light transport under the surface. FontCode alters the glyphs of each character continuously on a font manifold and uses font parameters to embed information. They provide encoding techniques that our system can take advantage of to communicate time-based information. 

AniCode uses a small printed code affixed to or embedded in an object and depends on a photograph of the object itself to drive the generation of a video for communication. Rather than using coded artifacts to redirect consumers to internet sites, we use the coded information to segment and animate a consumer-captured image to present visual information. After a one-time installation of an application on their mobile device, the consumer never has to access the internet to observe information provided by the author through coded artifacts affixed to objects in scenes. This improves robustness of the communication, because the consumer does not need network access that may be difficult in remote, congested or restricted areas, and the author does not have to ensure that a server providing videos is always available to provide information. This improves privacy (although does not guarantee it in any way), because neither the consumer's request nor the information they receive is transferred over the internet, and it requires people to be on-site to be able to view the animation, restricting those away from the scene from obtaining the information. 

%% file: 3_framework_overview.tex
\section{Framework Overview}
\label{sec:3}

The AniCode framework enables authors to specify an animation based on a captured image of physical artifacts. A tool guides the author through specifying the animation and generating a visual code containing the specification. The author places the code on an artifact to communicate with consumers. We use QR codes in our current implementation because tools to create them are readily available. To view an authored animation, the consumer captures an image of a scene that contains the coded artifact. Our mobile application generates a personalized animation on the fly without accessing the internet. The key pieces of our pipeline are: 

Author's side:
\begin{enumerate}
 \item The author takes an image of a static scene containing objects and a reference QR code as landmarks.
 \item AniCode performs image segmentation, and the author selects segments and specifies various animation information in the authoring interface.
 \item The author generates a new QR code based on the animation information and reference landmarks, then prints it to replace the reference QR code.
\end{enumerate}

Consumer's side:
\begin{enumerate}
 \item The consumer opens the mobile app with a smartphone, which directs the view to the scene with the authored QR code. The app takes a picture when the registration error is below a set threshold.
 \item The app performs image segmentation on the consumer's image. For each keyframe, it generates a mask of the region to be animated based on segment features matched with those in the author's image.
 \item The app generates the animation on the fly by using only the consumer's image and decoded information. The consumer views a visually contextualized demonstration privately.
\end{enumerate}

In the following sections we detail out methodology (Section \ref{sec:4}), show sample applications (Section \ref{sec:5}), describe a user study evaluating our system (Section \ref{sec:6}), and close with limitations and future work (Section \ref{sec:7}).

%% file: 4_methodology.tex
\section{Methodology}
\label{sec:4}

In this section, we elaborate on the design and implementation of 
the AniCode framework. First, we explore image segmentation algorithms to identify the regions to be animated. Then we show four animation schemes to convey the author's ideas -- 2D transformation, 3D transformation, color transformation, and annotation. We present an authoring interface in HTML and JavaScript that assists authors in converting an animation into a set of encoded numbers. Finally, we describe a network-free consumer-side mobile application on Android that can guide consumers to the correct view, take a picture, decode the authored information, and generate a personalized animation on the fly.

\subsection{Regions to Be Animated}
Authors need to specify a region of interest (ROI) to be animated in each keyframe. With a mask of an ROI can we apply image transformations. An ROI could be defined by a list of polygon vertices, but the desired ROI boundary may be in a somewhat different location in the consumer's image than in the author's. To make the ROI specification more robust, we use image segmentation.

The image segmentation 
should assign a segment $ID$ $(1\leq ID\leq K)$ to each pixel where $K$ is the number of segments in the image. Among many image segmentation algorithms are superpixel-based methods. After considering various options such as simple linear iterative clustering (SLIC) \cite{achanta2012slic} and superpixels extracted via energy-driven sampling (SEEDS) \cite{van2012seeds}, we selected linear spectral clustering (LSC) \cite{li2015superpixel} that produces compact and uniform superpixels with low computational cost. 

\begin{figure}[h]
\centerline{
\includegraphics[width=1.0\linewidth]{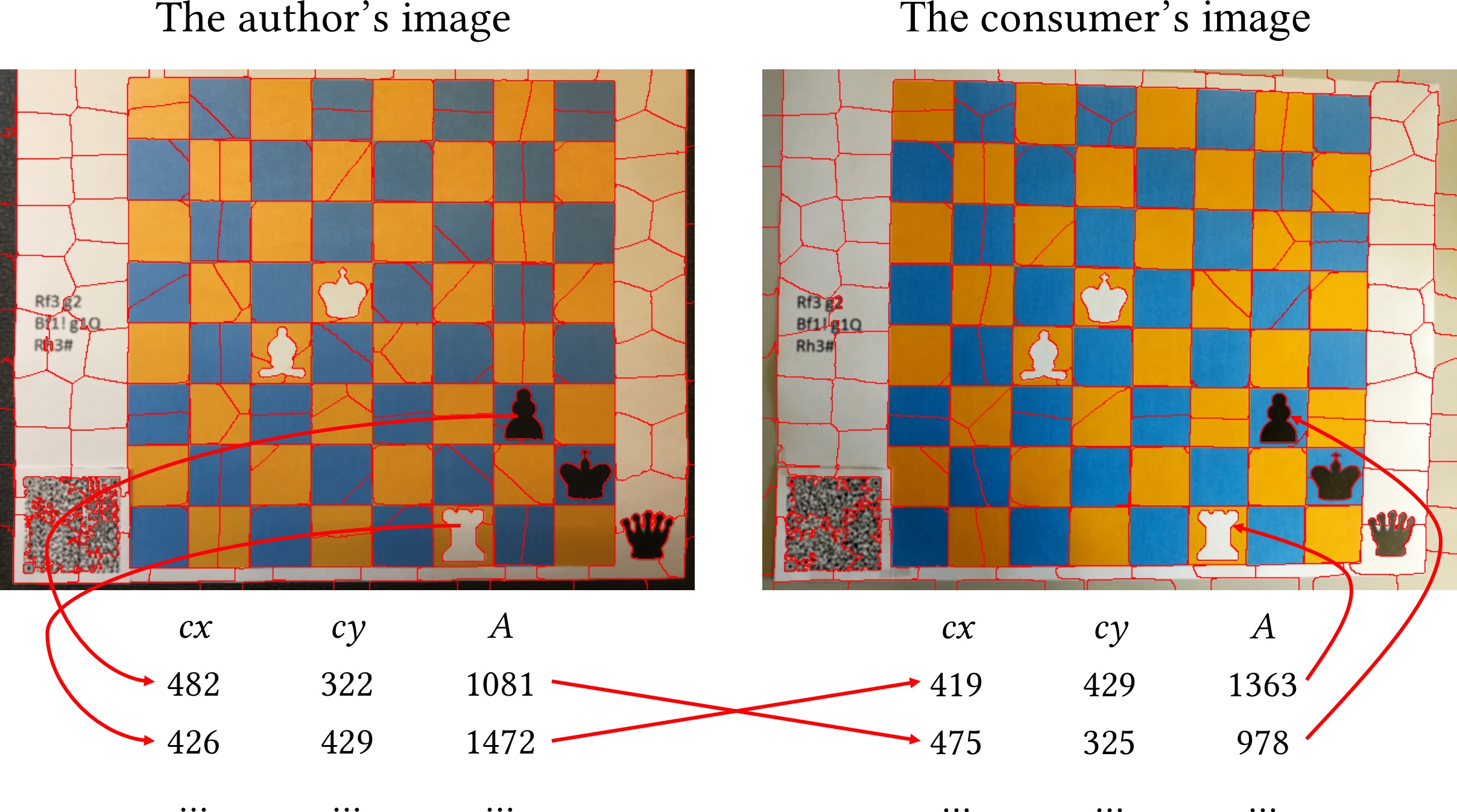}
}
\caption{Image segmentation results on both images using the LSC algorithm. The QR code stores segment features of the author's ROIs. The app generates masks of the consumer's ROIs based on feature matching.}
\label{fig:segmentation}
\end{figure}

In the LSC algorithm (applied in Fig.~\ref{fig:segmentation}) 
a pixel $\bm{p}=(l, \alpha, \beta, x, y)$ in the CIELAB color space (every component is linearly normalized to $[0,1]$) is mapped to a ten dimensional vector in the feature space. 
Taking as input the desired number of superpixels $K$, the algorithm uniformly samples $K$ seed pixels and iteratively performs weighted $K$-means clustering. 
It achieves local compactness by limiting the search space of each cluster. After the algorithm converges, it also enforces the connectivity of superpixels by merging small superpixels into neighboring bigger ones.

During the authoring process, the author chooses two parameters of the image segmentation algorithm (via sliders and interactive feedback): the average superpixel size and the superpixel compactness factor. Two other parameters are the number of iterations and the size of superpixels that should be absorbed into bigger ones. We fix them in our system since they have little effect on the result. Once the algorithm achieves a satisfactory result, it generates a matrix of segment IDs with the same size as the original image. For each keyframe to be animated, the author can select one or multiple segments as the ROI in the authoring interface. 
We extract segment features with a view to coding efficiency. We use three numbers to represent a segment, center coordinates $(cx, cy)$ and the area $A$ of the segment. The consumer-side app executes the LSC algorithm using the same parameters after the consumer takes a registered image. Before generating the animation, it generates an ROI mask for each keyframe based on matching the new segmentation result and the author's encoded features in terms of weighted mean squared error. In a mathematical representation, if the author selects $n$ segments for the ROI in the current keyframe, together with $3n$ features $cx'_i, cy'_i, A'_i$ where $1\leq i\leq n$, the matched ROI in the consumer's image is the union of all the matched segments:
$$
\mathrm{ROI} = \cup_{i=1}^{n} \mathrm{segment}_{ID_i}
$$
$$
ID_i = \mathop{\mathrm{argmin}}_j \quad (cx_j - cx'_i)^2 + (cy_j - cy'_i)^2 + \frac{1}{1000} (A_j - A'_i)^2
$$

\subsection{Animation Schemes}
\begin{figure}[h]
\includegraphics[width=0.32\linewidth]{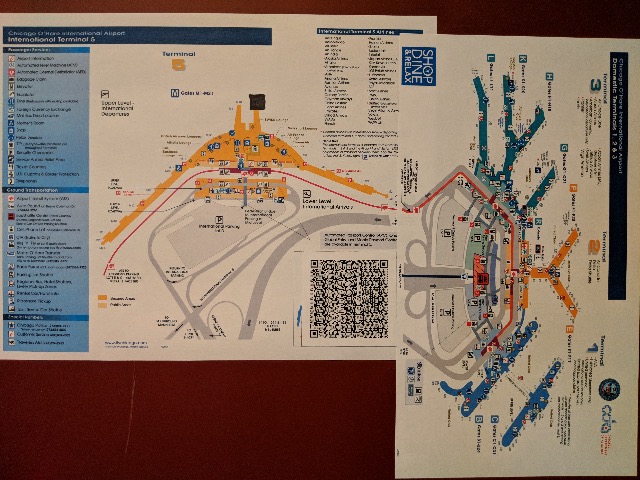}
\includegraphics[width=0.32\linewidth]{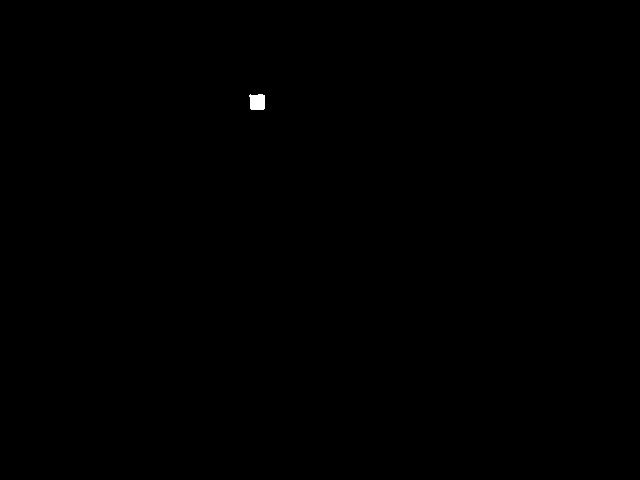}
\includegraphics[width=0.32\linewidth]{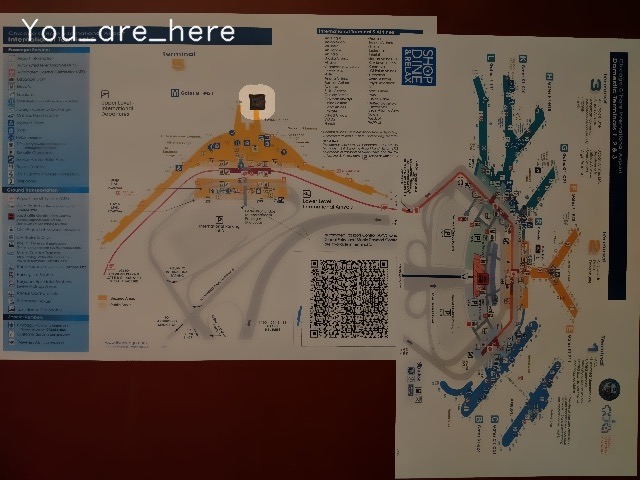}
\includegraphics[width=0.32\linewidth]{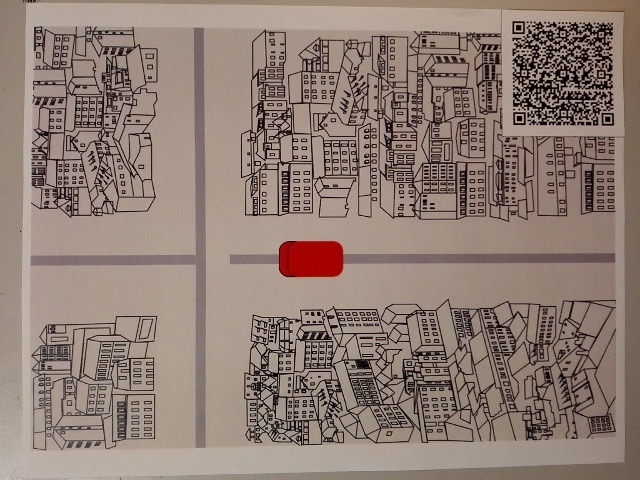}
\includegraphics[width=0.32\linewidth]{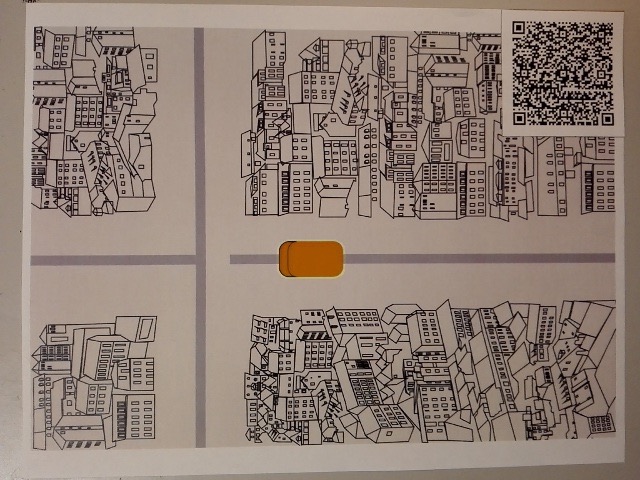}
\includegraphics[width=0.32\linewidth]{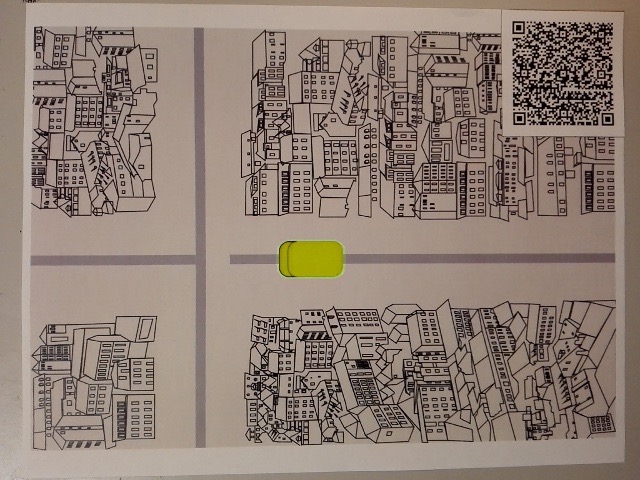}
\includegraphics[width=0.32\linewidth]{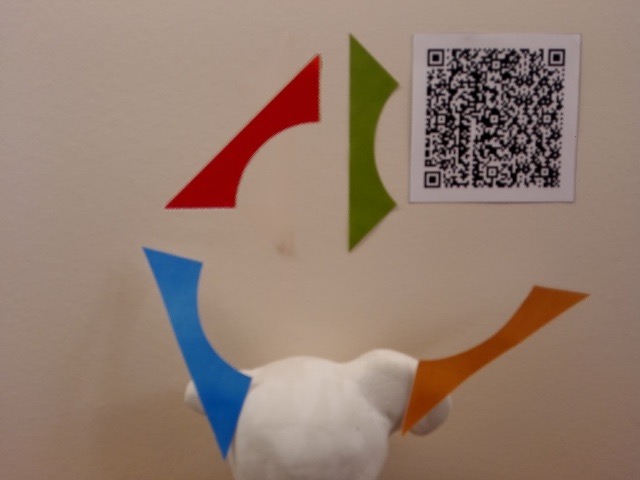}
\includegraphics[width=0.32\linewidth]{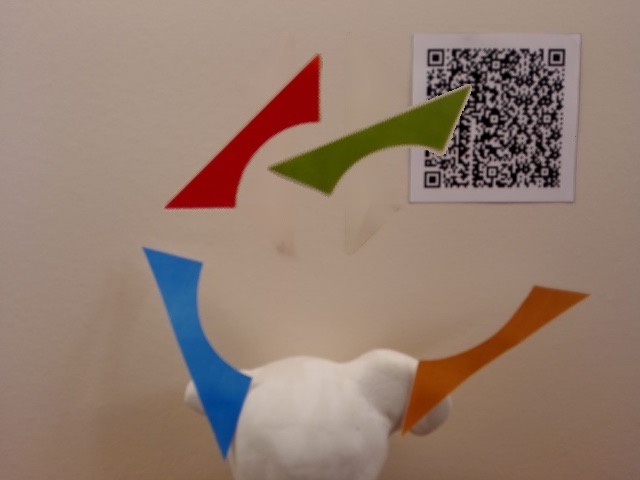}
\includegraphics[width=0.32\linewidth]{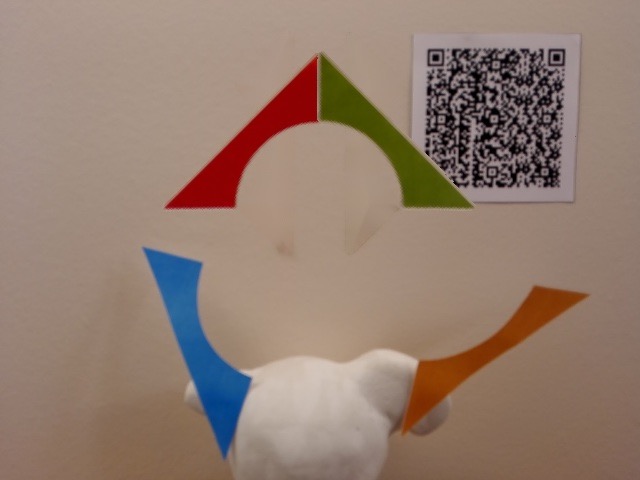}
\includegraphics[width=0.32\linewidth]{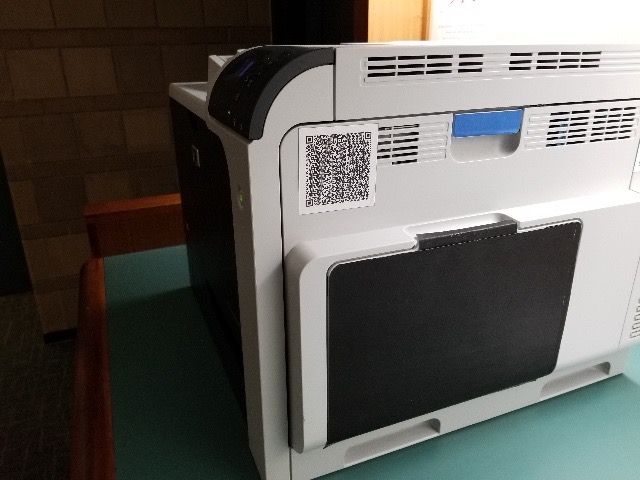}
\includegraphics[width=0.32\linewidth]{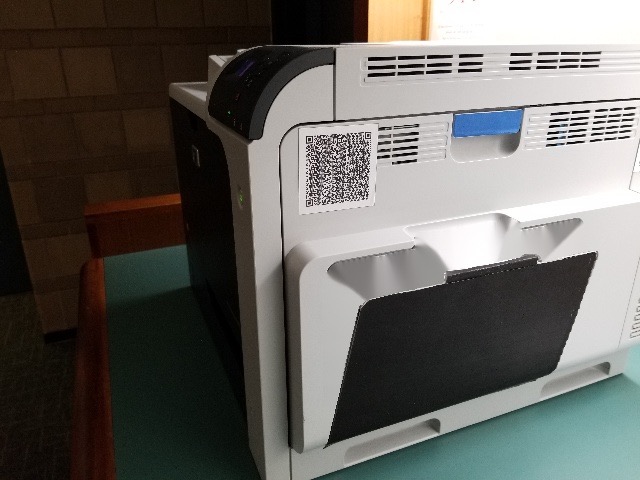}
\includegraphics[width=0.32\linewidth]{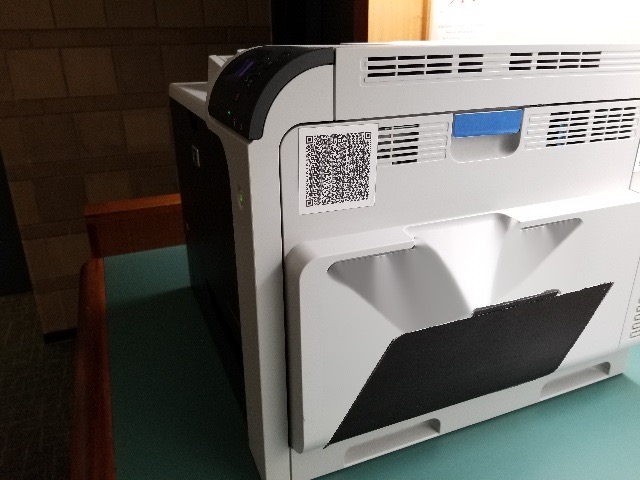}
\caption{Animation schemes. From left to right in the first row: original image, ROI mask, and annotation frame. Second row: color transformation frames. Third row: 2D transformation frames. Last row: 3D transformation frames.}
\label{fig:schemes}
\end{figure}

To create an animation, the author selects a region to be animated and applies a transformation to it in each keyframe. 
Common operations include 2D geometric transformation, 3D geometric transformation, color transformation, and annotation. We design four animation schemes (Fig.~\ref{fig:schemes}). All the animation schemes except annotation can use linear or quadratic interpolation during the animation to generate changes at different temporal rates.

\textbf{2D Transformation}. It is basic to use a 2D rigid transformation to describe object motion on the image plane. The author can specify translation $t_x$, $t_y$ 
as well as rotation $\theta$ after selecting an ROI. We assume the rotation is around the ROI center $(c_x, c_y)$ since it is more intuitive to authors. The author needs to specify the duration $T$ of the current keyframe in seconds and linear or quadratic interpolation. Assuming 30 frames per second (fps) in the video, we apply the following geometric transformation to the selected ROI in the $i^{th}$ frame:
$$
M=
\begin{bmatrix}
1 & 0 & r t_x \\
0 & 1 & r t_y \\
0 & 0 & 1
\end{bmatrix}
\begin{bmatrix}
1 & 0 & c_x \\
0 & 1 & c_y \\
0 & 0 & 1
\end{bmatrix}
\begin{bmatrix}
\cos{r \theta} & \sin{r \theta} & 0 \\
-\sin{r \theta} & \cos{r \theta} & 0 \\
0 & 0 & 1
\end{bmatrix}
\begin{bmatrix}
1 & 0 & -c_x \\
0 & 1 & -c_y \\
0 & 0 & 1
\end{bmatrix}
$$
where $r$ is the interpolation ratio. $r=\frac{i}{30T}$ for linear interpolation and $r={\frac{i}{30T}}^2$ for quadratic interpolation. After the current ROI is moved to a different location, there may be a hole at the original location. 
We use the Telea algorithm \cite{telea2004image} for image inpainting that takes the original image and ROI mask as input before generating the animated frames. We also dilate the ROI mask using an elliptical structuring element of size 2 to further improve the visual effect.

\textbf{3D Transformation}. When authors want to describe object motion in 3D space, a 2D rigid transformation is not sufficient. However, it is not practical to store any 3D mesh information in a QR code. Therefore, we use a subspace of 3D transformation, which is 2D perspective transformation on the image plane, because it can describe the motion in 3D space. Instead of translation and rotation, the author needs to specify four pairs of corresponding points $\bm{p}_j=(x_j,y_j)$, $\bm{p}'_j=(x'_j,y'_j)$ before and after the transformation to obtain the homography. Given the duration $T$ in seconds, 30 fps, in the $i^{th}$ frame, we can solve the transformation matrix $M$ using a linear system solver as follows:
$$
w_j
\begin{bmatrix}
x_j + r(x'_j-x_j) \\
y_j + r(y'_j-y_j) \\
1
\end{bmatrix}
=
\begin{bmatrix}
m_{xx} & m_{xy} & m_{xw} \\
m_{yx} & m_{yy} & m_{yw} \\
m_{wx} & m_{wy} & 1 \\
\end{bmatrix}
\begin{bmatrix}
x_j \\
y_j \\
1
\end{bmatrix},
j=1,2,3,4
$$
where the interpolation ratio $r=\frac{i}{30T}$ for linear interpolation and $r={\frac{i}{30T}}^2$ for quadratic interpolation. $w_j$ is a normalizer for homogenous coordinates. 

\textbf{Color Transformation}. Color transformation can help emphasize a certain region in the image or produce artistic effects. To preserve textures on the object, we convert the image from RGB space to HSV space and change the hue channel uniformly given $\Delta h$ from the author. 
In the $i^{th}$ frame, we change the RGB values of pixels within the ROI mask (dilated using an elliptical structuring element of size 2) as follows, where we use the same interpolation ratio $r$ defined in previous animation schemes:
$$
\begin{bmatrix}
R' \\
G' \\
B'
\end{bmatrix}
=
HSV2RGB
\begin{pmatrix}
RGB2HSV
\begin{pmatrix}
\begin{bmatrix}
R \\
G \\
B
\end{bmatrix}
\end{pmatrix}
+
\begin{bmatrix}
r \Delta h \\
0 \\
0
\end{bmatrix}
\end{pmatrix}
$$

\textbf{Annotation}. In expository videos, it is useful to provide textual annotation to explain the functionality. A QR code is capable of encoding some short text for an ROI. Given this information, we dilate the ROI mask using an elliptical structuring element of size 10 and perform bilateral filtering on the background to focus the attention of consumers based on psychological models \cite{yeshurun1998attention}:
$$
\tilde I (\bm{p}) = 0.7 \frac{\sum_{\bm{q}\in \Omega} I(\bm{q})\cdot f(||I(\bm{q})-I(\bm{p})||)\cdot g(||\bm{q}-\bm{p}||)}{\sum_{\bm{q}\in \Omega} f(||I(\bm{q})-I(\bm{p})||) \cdot g(||\bm{q}-\bm{p}||)}, \forall \bm{p} \notin ROI
$$
where $\bm{p}$ is a pixel in the image, $\Omega$ is the window of diameter 15 centered in $\bm{p}$, $f$ is the range kernel for smoothing differences in intensities with a sigma of 80, and $g$ is the spatial kernel for smoothing differences in coordinates with a sigma of 80. Note that we also reduce the intensity of the non-ROI region by 30\% so that the ROI can stand out. The text is presented at the top left corner of the filtered image. Annotation also works well with color transformation if the author wants to emphasize an ROI.

\subsection{Authoring Interface}
\begin{figure}[h]
\centerline{
\includegraphics[width=1.0\linewidth]{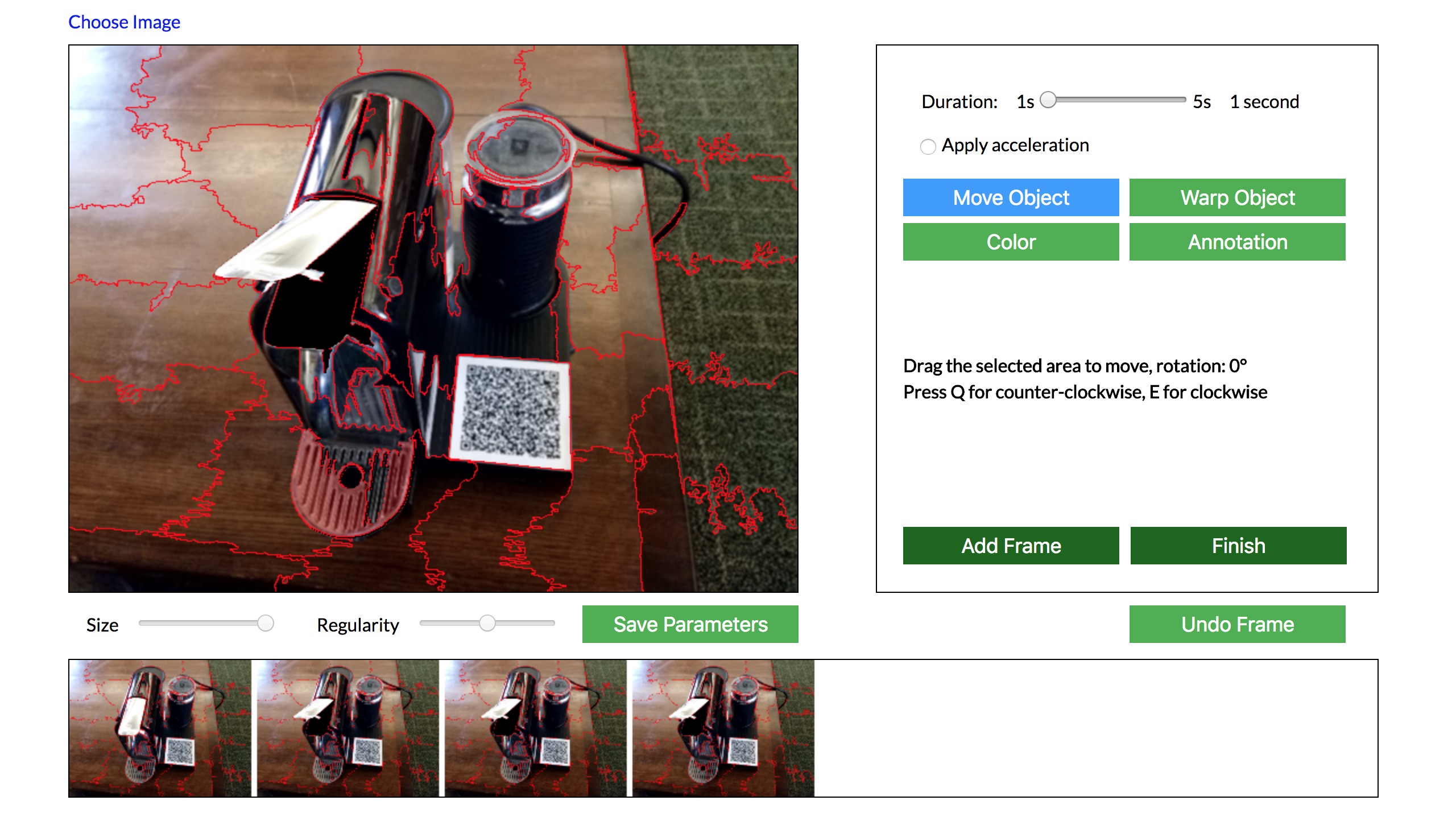}
}
\caption{Authoring interface. The author can upload an image and then select segments to form an ROI for each keyframe. The author also selects an animation scheme in each keyframe and interactively specifies information to be encoded. The canvas is updated in real time. Our system generates an animation preview and a QR code when authoring is finished.}
\label{fig:interface}
\end{figure}

After obtaining the image to animate and its segmentation, the author can proceed to add transformations to the scene using the authoring interface. In our current system, authoring is performed on a laptop or desktop machine. The interface (Fig.~\ref{fig:interface}) requires the author's image and the set of files resulting from the segmentation program, which contains a matrix of segment IDs and a table of segment features. When an image is loaded, the system detects the reference QR code and stores the landmark coordinates. The author can tune the two segmentation parameters by adjusting sliders and visualizing the segmentation result. 

The author can apply our animation schemes to a selected ROI in the image. Each time the author clicks on a segment, it is highlighted in red. All the highlighted segments add up to an ROI for the current keyframe, for which the duration can be indicated using a slider in the interface. After the author selects an animation scheme from the right panel, corresponding instructions appear. Here we use the terms ``Move Object'' and ``Warp Object'' instead of ``2D/3D transformation'' to appeal to a wider audience of users. For 2D transformation, the author can drag the ROI to a new position and rotate 
For 3D transformation, the author can drag four green pins to the corners of the ROI, and drag four purple pins to new positions in order to warp the ROI. When the purple pins are dragged, the interface computes a homography and warps the ROI in real time. For color transformation, the author can use a slider with a hue spectrum above it. 
The interface initializes the value of the slider to average hue of the current ROI, and updates it in real time to reflect color change. For annotation, the author can simply type in the desired text. After the addition of each keyframe, the interface updates the canvas and matrix of segment IDs. A thumbnail of the current canvas is rendered at the bottom as well so the author can keep track of all the keyframes. The author can also undo a keyframe if not satisfied with the previous frame. 

When authoring is finished, the system generates a preview animation based on the author's image and a new QR code 
Since the QR code stores a limited amount of information, the author is alerted if the amount of information to be encoded has exceeded the maximum allowed. As a tradeoff between capacity and size, the version 13 QR code that we use can encode 483 alphanumeric characters (about seven keyframes) with the data correction level M. The new QR code uses the following encoding: 
\begin{itemize}
\item 8 numbers, $xy$-coordinates of four landmarks of the reference QR code.
\item 4 numbers, parameters of the LSC algorithm, i.e., the average superpixel size, the superpixel compactness factor, the size of superpixels that should be absorbed into bigger ones, and the number of iterations. Note that only the first two parameters can be tuned in the interface, but we reserve all the fields in case the interface should be updated.
\item An integer $m$, the number of keyframes to be animated. For each of the following $m$ lines
\begin{itemize}
  \item An integer $n$, indicating how many segments there are in the current ROI.
  \item Following are $3n$ integers for features of each segment, i.e., $xy$-coordinates of the segment center and the segment area.
  \item An integer $\tau$ to represent the animation type of the current keyframe.
  \begin{itemize}
    \item If $\tau=0$ or $\tau=4$, this type is 2D transformation. Following are 4 numbers, i.e., horizontal translation, vertical translation, rotation, and duration for each 2D transformation. Linear interpolation is used during the animation when $\tau=0$ and quadratic one is used when $\tau=4$.
    \item If $\tau=1$ or $\tau=5$, this type is 3D transformation. Following are 16 numbers, i.e., $xy$-coordinates of four pairs of corresponding points before and after the perspective transformation. Another number is also needed for duration. Linear interpolation is used during the animation when $\tau=1$ and quadratic one is used when $\tau=5$.
    \item If $\tau=2$ or $\tau=6$, this type is color transformation. Following are 2 numbers, i.e., change of the hue channel and duration in seconds. Linear interpolation is used during the animation when $\tau=2$ and quadratic one is used when $\tau=6$.
    \item If $\tau=3$, this type is annotation. Following are a string and a number, i.e., text content of the annotation and duration in seconds.
  \end{itemize}
\end{itemize}
\end{itemize}

\subsection{Consumer-Side Application}
\begin{figure}[b]
\includegraphics[width=0.7\linewidth]{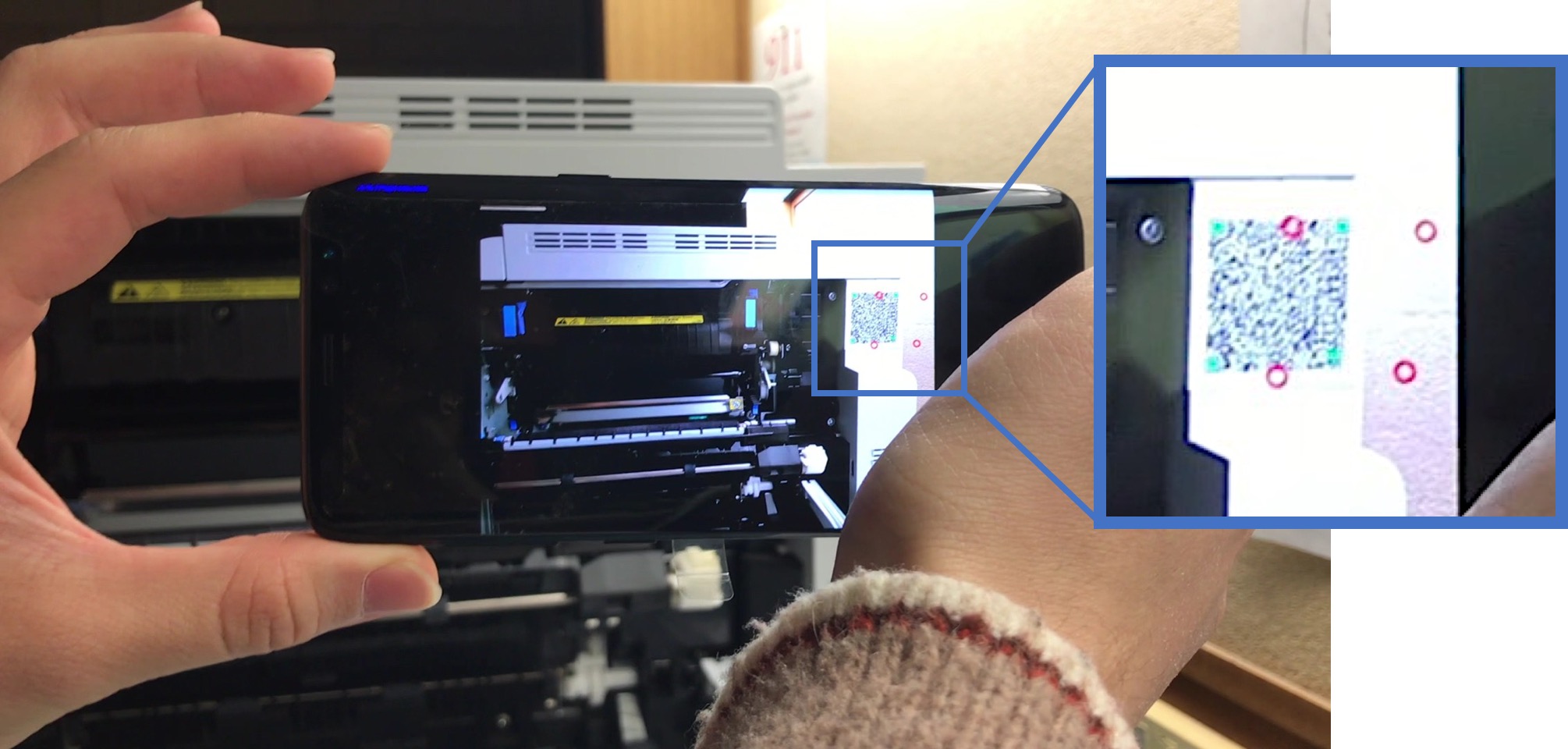}
\caption{Consumer-side application. Red circles are where the landmarks are supposed to be, and green dots are the current landmarks detected in real time. The animation starts once the view is registered.}
\label{fig:app}
\end{figure}

The consumer-side application (Fig.~\ref{fig:app}) in the AniCode framework works on a handheld mobile device.
When the consumer holds a smartphone towards a scene that has authored objects, the app automatically detects and decodes the QR code in the scene using the ZXing QR code framework \cite{zxing}. The first piece of information is the landmark coordinates of the reference QR code in the author's image. Four static red circles are rendered at these locations on the screen. The app also draws four green dots on the screen in real time at the landmark coordinates of the current QR code based on the detection result. For a smoother user experience, we also limit the QR code detection window to a narrower area instead of the whole image based on landmark coordinates of the reference QR code. The consumer registers the view to the author's specification by aligning the dots with the circles. Once the error is below a set threshold, the app takes a image and writes it to the external storage in order to generate the animation.

The app automatically switches to the animation generation mode as soon as the image is registered. First, the rest of the decoded information is processed. It performs the LSC image segmentation algorithm using the same parameters on the consumer's image downscaled to $640 \times 480$ to achieve a good runtime performance without compromising much visual quality. The app generates an ROI mask for each keyframe based on matching the features of segments with the features stored in the QR code. It is worth noting that a segment may be moved to somewhere else in previous 2D or 3D transformation keyframes before animating the current keyframe. Therefore, we maintain a transformation stack for each segment to keep track of their locations. After segment matching is complete, we apply the accumulated transformation to each segment to obtain the intended ROI mask. Once we have a ROI mask for each keyframe, it is easy to parse other parameters in this keyframe to generate the image sequence using our animation schemes. Our current implementation is an Android app using OpenCV in Java \cite{opencv}. 

%% file: 5_sample_applications.tex
\section{Sample Applications}
\label{sec:5}
We demonstrate sample applications 
in the following three areas: expository media, cultural heritage and education, creative art and design. We use static figures to summarize the examples; the animations themselves are available in our supplementary video. Our supplementary portfolio includes more examples and their detailed descriptions.

\subsection{How do I use this?}
\begin{figure}[b]
\includegraphics[width=0.44\linewidth]{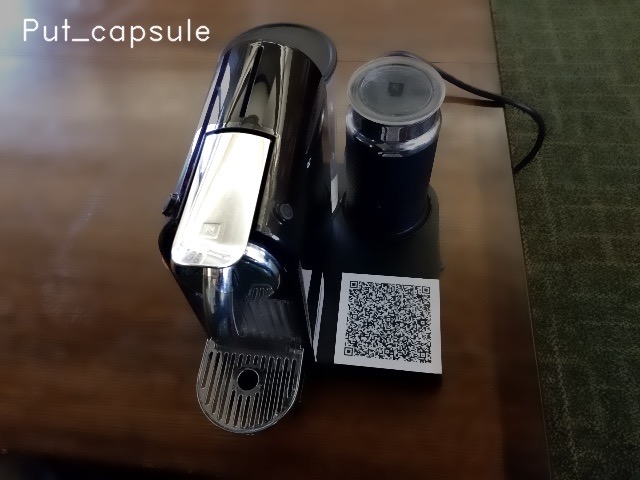}
\includegraphics[width=0.44\linewidth]{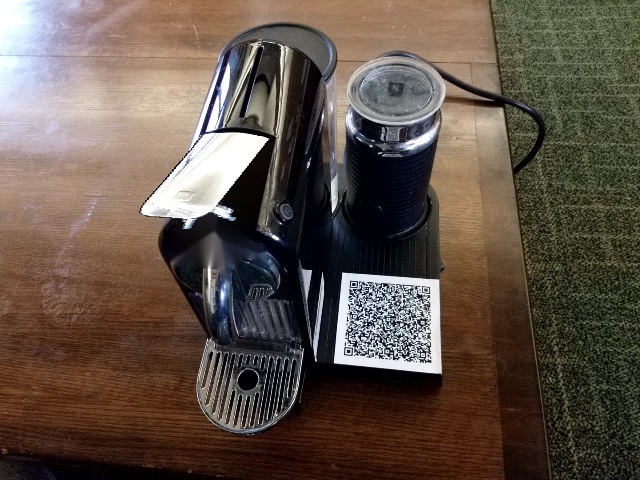}
\includegraphics[width=0.44\linewidth]{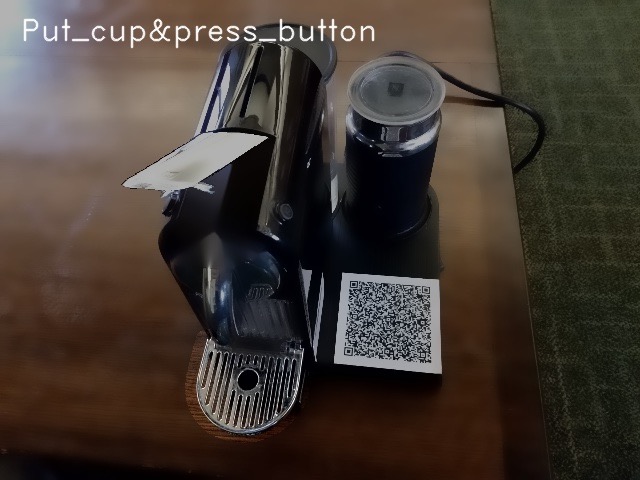}
\includegraphics[width=0.44\linewidth]{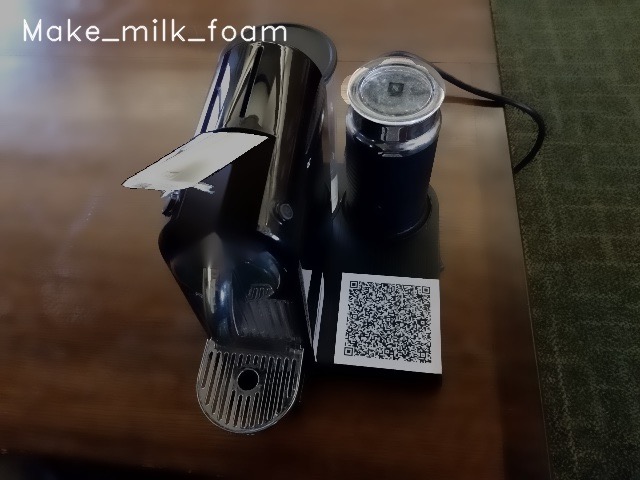}
\includegraphics[width=0.44\linewidth]{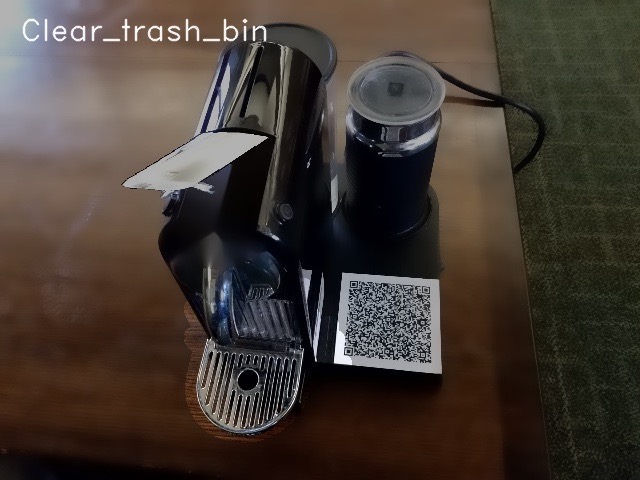}
\includegraphics[width=0.44\linewidth]{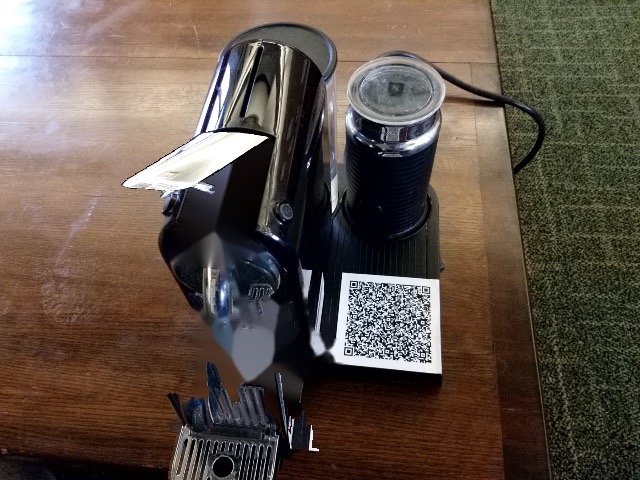}
\caption{Coffee maker. This animation communicates the operations of lifting the lid, placing the cup, making milk foam, and clearing the trash bin. {\em Animation in supplemental video.}}
\label{fig:coffee}
\end{figure}
\begin{figure}[b]
\includegraphics[width=0.44\linewidth]{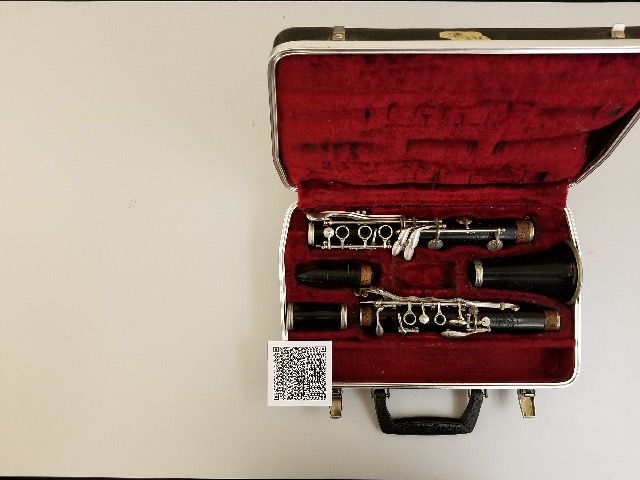}
\includegraphics[width=0.44\linewidth]{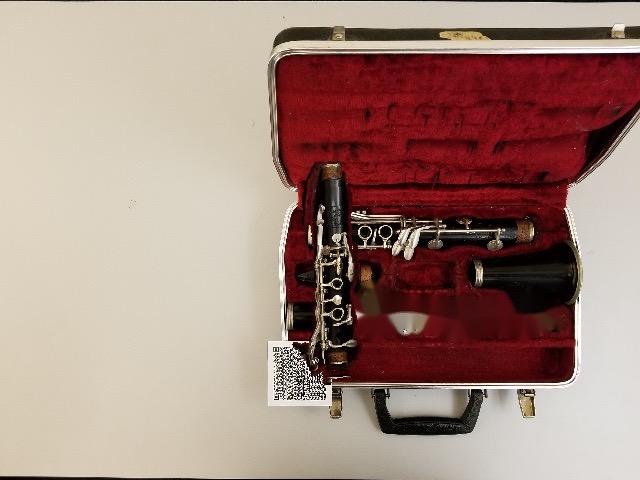}
\includegraphics[width=0.44\linewidth]{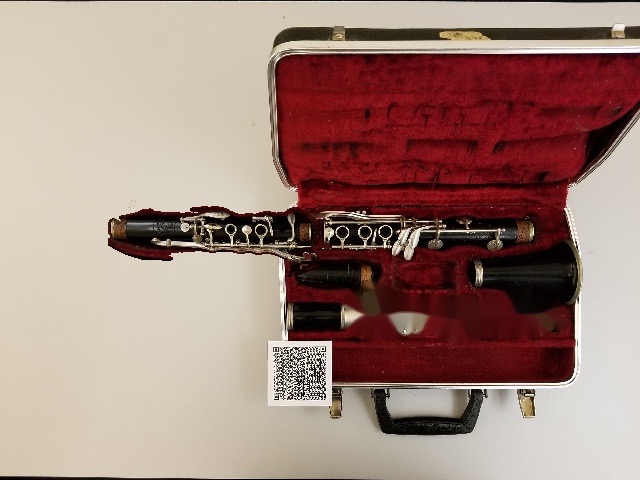}
\includegraphics[width=0.44\linewidth]{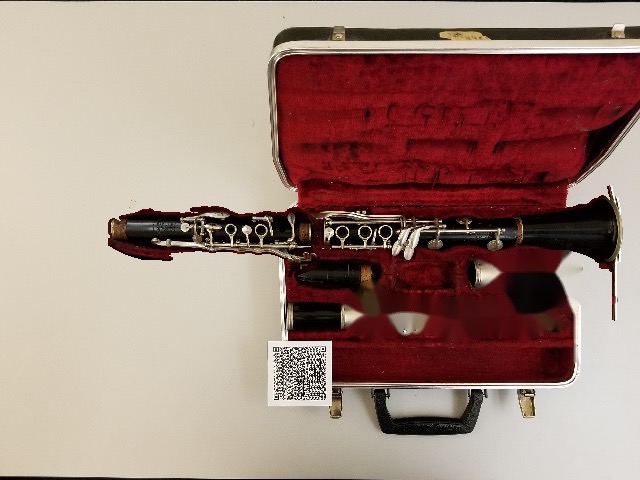}
\includegraphics[width=0.44\linewidth]{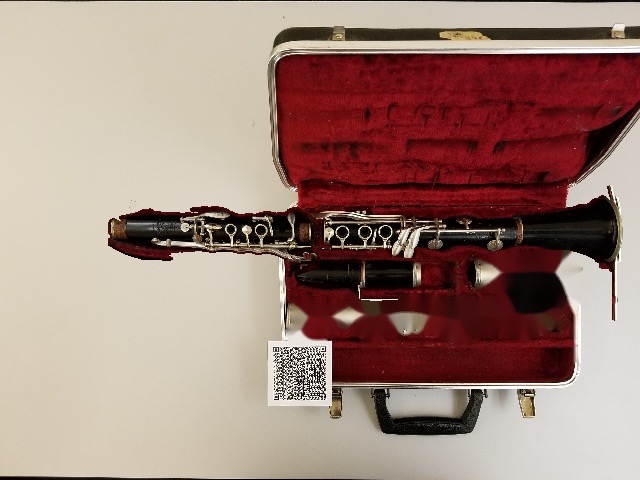}
\includegraphics[width=0.44\linewidth]{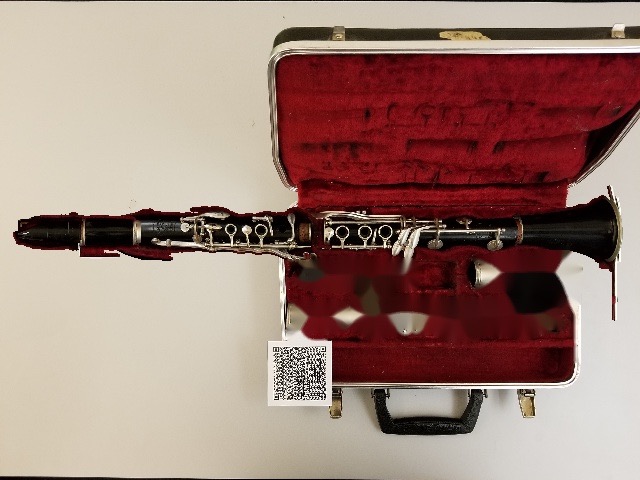}
\caption{Assembly of a clarinet. The upper tube flips over to connect to the lower tube. The mouthpiece, barrel, and bell move to their corresponding locations to show the assembly. {\em Animation in supplemental video.}}
\label{fig:clarinet}
\end{figure}
\begin{figure}[h]
\includegraphics[width=0.44\linewidth]{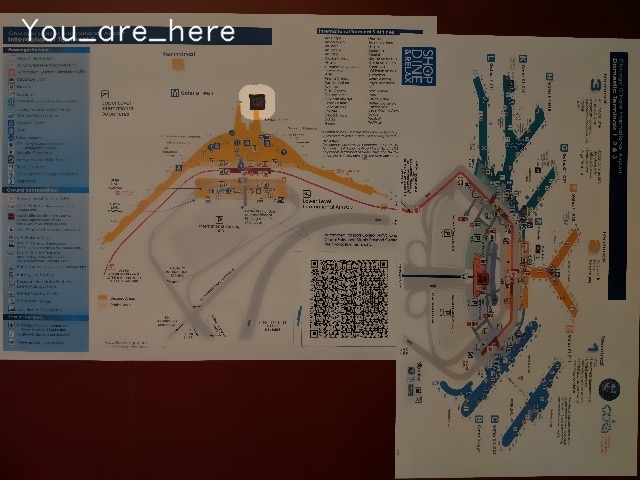}
\includegraphics[width=0.44\linewidth]{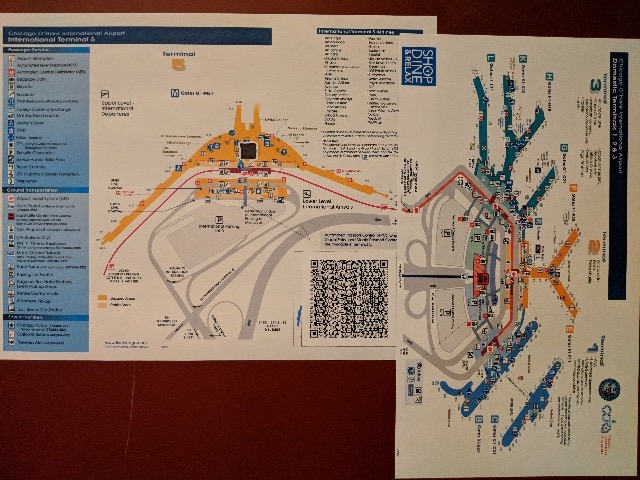}
\includegraphics[width=0.44\linewidth]{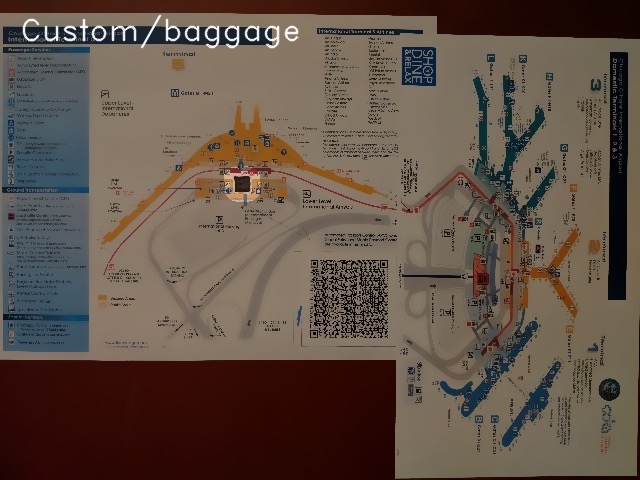}
\includegraphics[width=0.44\linewidth]{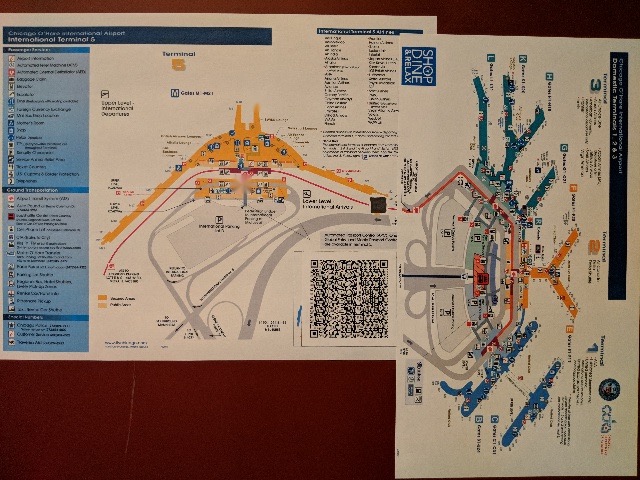}
\includegraphics[width=0.44\linewidth]{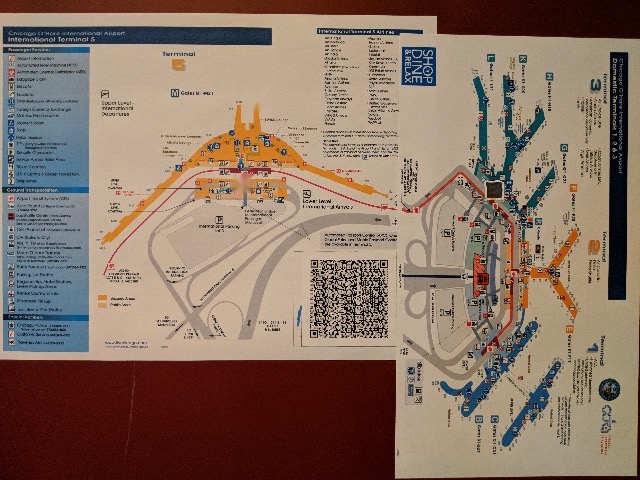}
\includegraphics[width=0.44\linewidth]{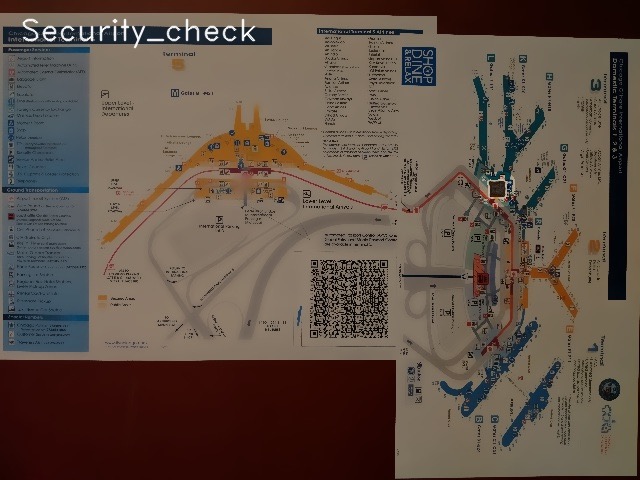}
\caption{Navigation in an airport. International arrivals need to pass through customs, board the airport shuttle to change terminals, and go through security again for domestic flights. {\em Animation in supplemental video.}}
\label{fig:map}
\end{figure}
A common problem is illustrating how to use a product. Products are designed, to the extent possible, with well-known visual affordances (e.g., the handle on a cup). Simplified diagrams are also sometimes printed on objects or booklets. Online videos describe and illustrate usage. All of these approaches have potential problems; an obvious visual affordance may not be possible to incorporate into the design, people may not be able to understand simplified diagrams, booklets may be lost, and online videos require network access and may show the product in a different view or lighting which make it difficult for users to relate to their particular situation. 

The AniCode framework makes it possible to show the consumer an animated video on how to use a product based on the consumer's current visual context. 
A specific example is shown in Fig.~\ref{fig:coffee}. The consumer needs to understand that they should lift the lid, put in a capsule, and put a cup under the nozzle to make coffee. They also need to know how to add water, clear the trash bin, and make milk foam. The author places a generated QR code on the coffee maker in advance, and the consumer views the animation generated on the fly in the mobile app. The author describes the lid being lifted using a 3D transformation keyframe, and the trash bin being pulled out using a 2D transformation keyframe. Annotations are also shown to indicate where to put the cup and make milk foam.

A second example is to assemble parts of an object. 
In Fig.~\ref{fig:clarinet}, we show a simple example putting together the sections of clarinet. The user is able to view a personalized video using the authored QR code. The pieces of the clarinet are moved to their appropriate locations using 2D transformation keyframes to show the final assembly of the instrument.

One of the strengths of our system, offline data, can be extremely useful in navigation. When traveling, people are often in situations without internet access to give guidance. When travelers arrive at an airport to make a connecting flight, they are often unfamiliar with the complex layout of the airport. 
In the example in Fig.~\ref{fig:map}, a passenger from an international flight has arrived at Chicago O'Hare International Airport and needs to pass through customs, board the airport shuttle to change terminals, and go through security for domestic flights.
Using our system, we can provide guidance for how to proceed to a specific set of gates. The animation shows the path to customs, then boarding of the shuttle, followed by the stop to leave the shuttle and pass through security.
The animation highlights the route that could be difficult to follow on a static map. 

Another important aspect to note is that consumers can obtain visual information without having to disclose their personal information using AniCode, whereas sending a request for navigation assistance over the internet broadcasts a consumer's location and the information requested. AniCode also requires people to be on-site to be able to view the animation, preventing those who are not allowed to enter the scene from obtaining the information. 

\subsection{Cultural Heritage and Education}
\begin{figure}[h]
\includegraphics[width=0.44\linewidth]{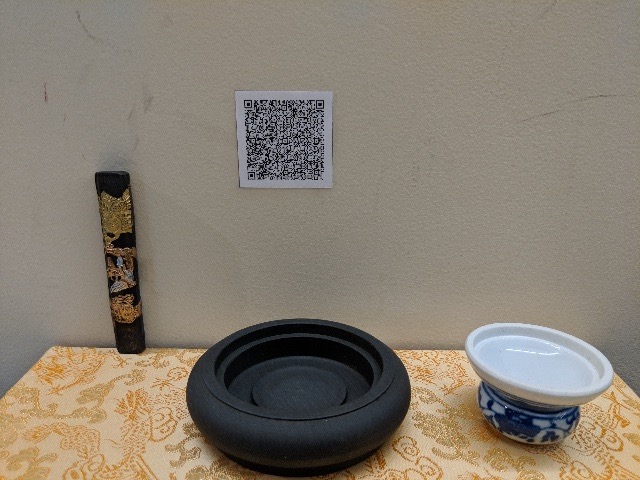}
\includegraphics[width=0.44\linewidth]{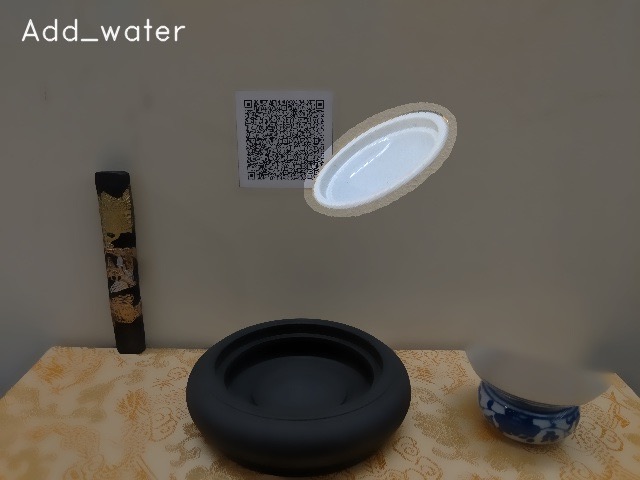}
\includegraphics[width=0.44\linewidth]{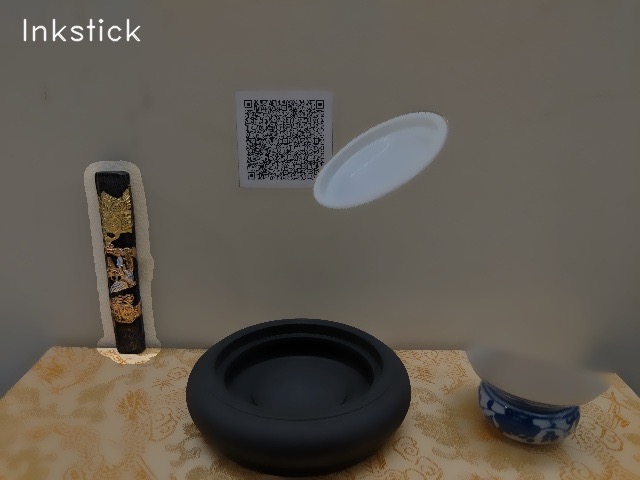}
\includegraphics[width=0.44\linewidth]{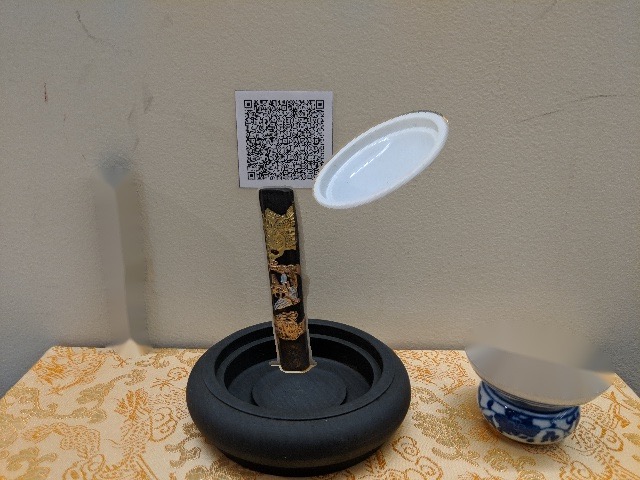}
\includegraphics[width=0.44\linewidth]{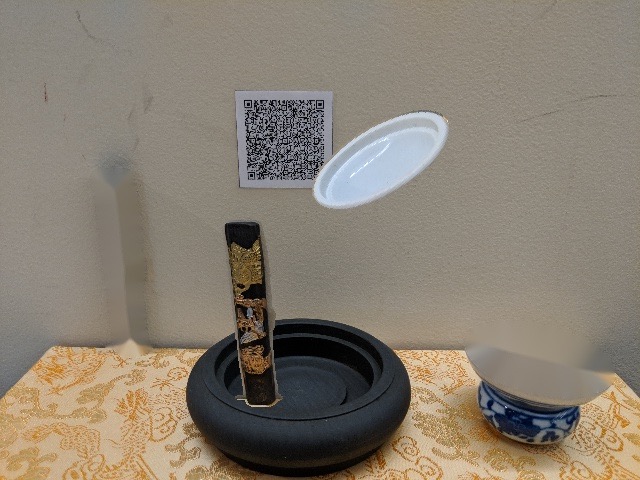}
\includegraphics[width=0.44\linewidth]{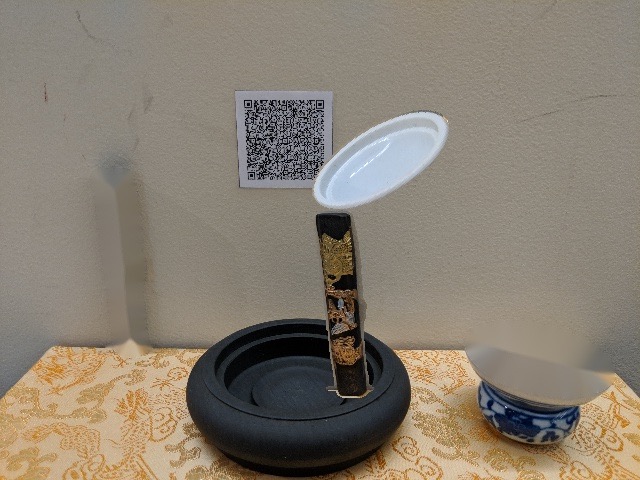}
\caption{Ink grinding. First, water is added to the inkstone. Then, the inkstick is highlighted with annotation and moved over the inkstone to show the process of grinding ink. {\em Animation in supplemental video.}}
\label{fig:ink}
\end{figure}

Most cultural heritage artifacts on display at museums and historical sites cannot be touched by visitors; their purpose is usually explained through text. However, text does not always effectively convey concepts that are inherently foreign to visitors. This is especially true when there are multiple artifacts involved, which are used in a complex manner. Providing information through kiosks, Wi-Fi, and cloud-based storage can be financially infeasible for small museums and remote sites. In such cases, visual communication is a more effective way to communicate knowledge to the audience, and showing the animation on the consumer's smartphone without accessing the internet gives the added bonus of a private and personalized experience.

An example in the heritage domain is the process of grinding ink in Fig.~\ref{fig:ink}, which was part of the standard method of writing in ancient Asia. First, the water container is lifted and moved over the inkstone to show the addition of water. Then, an annotation is displayed identifying the inkstick. Finally, the inkstick is moved to the location of the inkstone and ground over its surface to demonstrate ink production. 
\begin{figure}[h]
\includegraphics[width=0.44\linewidth]{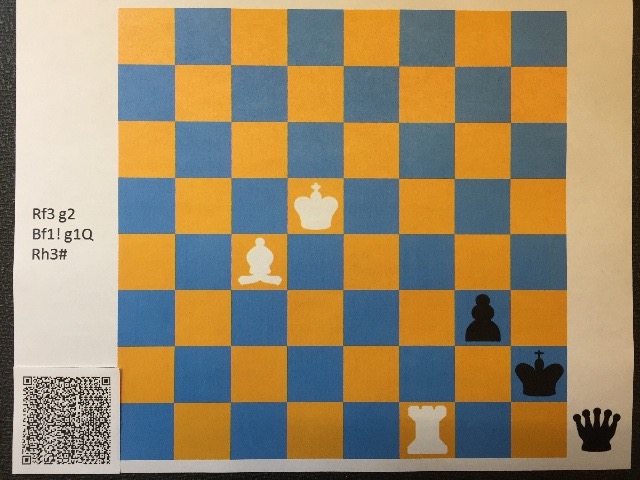}
\includegraphics[width=0.44\linewidth]{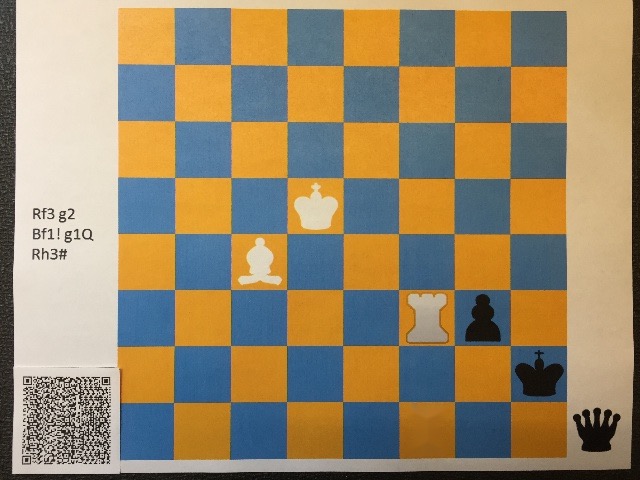}
\includegraphics[width=0.44\linewidth]{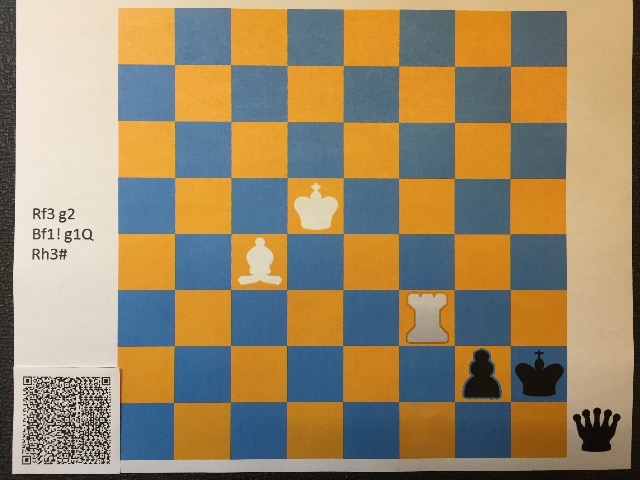}
\includegraphics[width=0.44\linewidth]{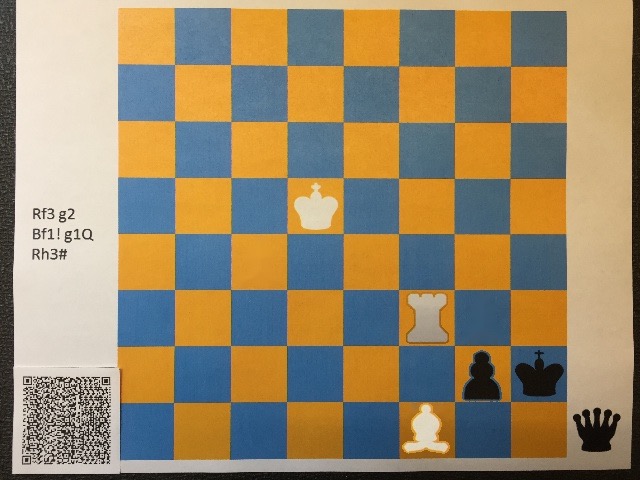}
\includegraphics[width=0.44\linewidth]{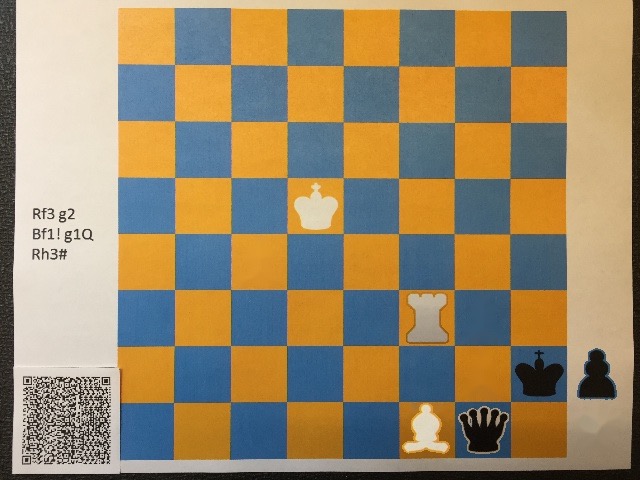}
\includegraphics[width=0.44\linewidth]{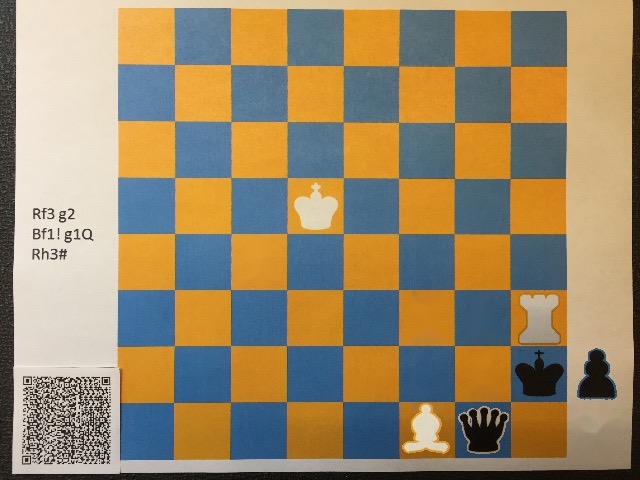}
\caption{A chess endgame. The moves can be recorded using the algebraic notation ``Rf3 g2, Bf1! g1Q, Rh3\#'', but it is far less engaging than an animation, especially for amateur players. {\em Animation in supplemental video.}}
\label{fig:chess}
\end{figure}

It is also possible to apply our system to education. We show a chess endgame in Fig.~\ref{fig:chess}. In chess training, it is important to understand possible moves and tactics in an endgame. A winning strategy for the white side is as follows. In the first round, the white rook moves to f3, and the black pawn moves to g2. In the second round, the white bishop makes a key move to f1, and the black pawn moves to g1 and is promoted to a queen. Finally, the black side is checkmated if the white rook moves to h3. Compared with the algebraic notation such as ``Rf3 g2, Bf1! g1Q, Rh3\#'', our system offers an interactive animation to convey ideas through a static poster.

\subsection{Creative Art and Design}
\begin{figure}[h]
\includegraphics[width=0.44\linewidth]{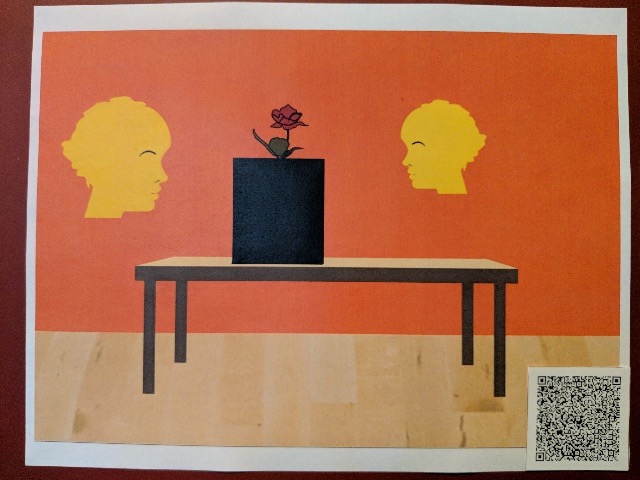}
\includegraphics[width=0.44\linewidth]{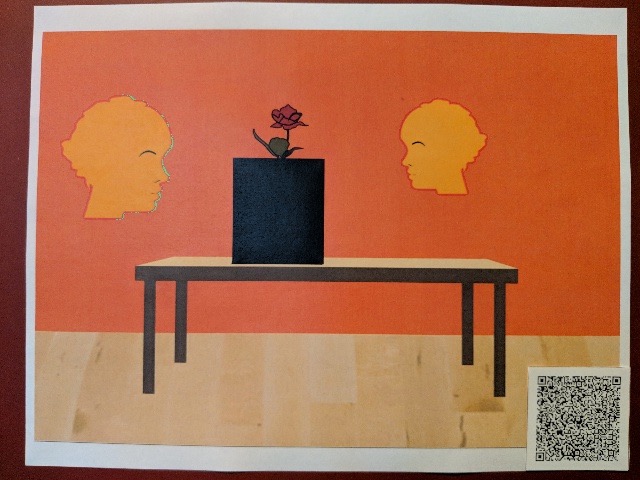}
\includegraphics[width=0.44\linewidth]{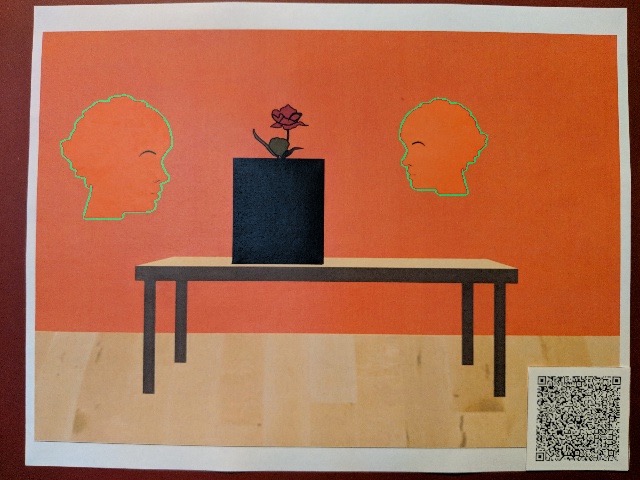}
\includegraphics[width=0.44\linewidth]{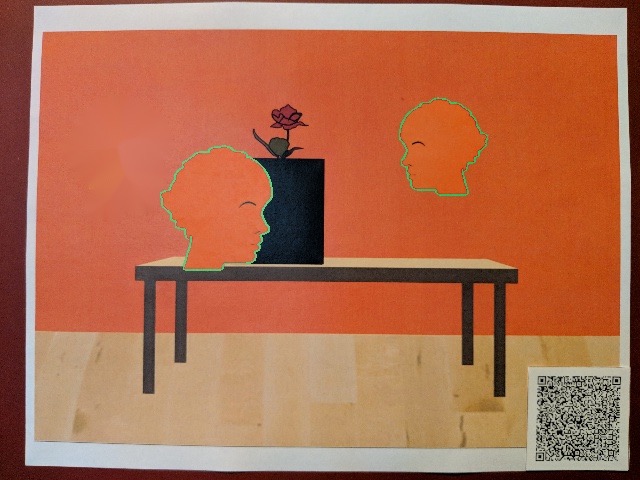}
\includegraphics[width=0.44\linewidth]{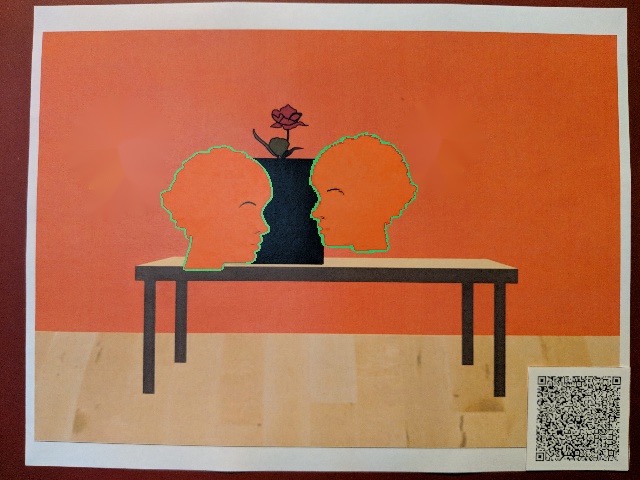}
\includegraphics[width=0.44\linewidth]{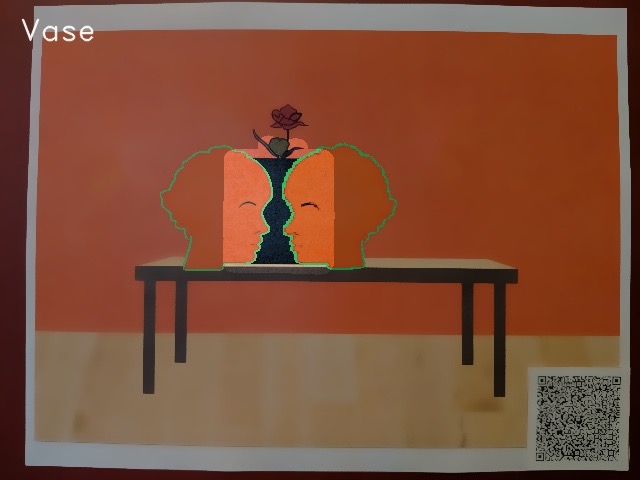}
\caption{The Rubin vase. Two yellow heads gradually change color to the background red, and a vase is carved out from the black box by the warped heads. {\em Animation in supplemental video.}}
\label{fig:vase}
\end{figure}

Artists can design animated posters using our system. For example, the poster in Fig.~\ref{fig:vase} initially shows two people appreciating a flower in a black box. The artist changes the color of the heads to merge them into the background, and warps them in such a way that a vase is carved out from the black box using keyframes of color and perspective transformations. With an authored QR code from our system, the consumer is able to understand the ideas that previously could not be conveyed through a static poster. 
Educators can also use this type of example to show the psychological concept of bistable images intuitively.
\begin{figure}[h]
\includegraphics[width=0.44\linewidth]{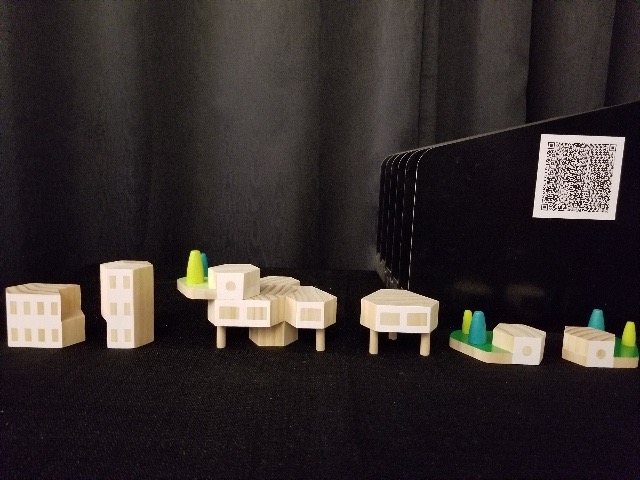}
\includegraphics[width=0.44\linewidth]{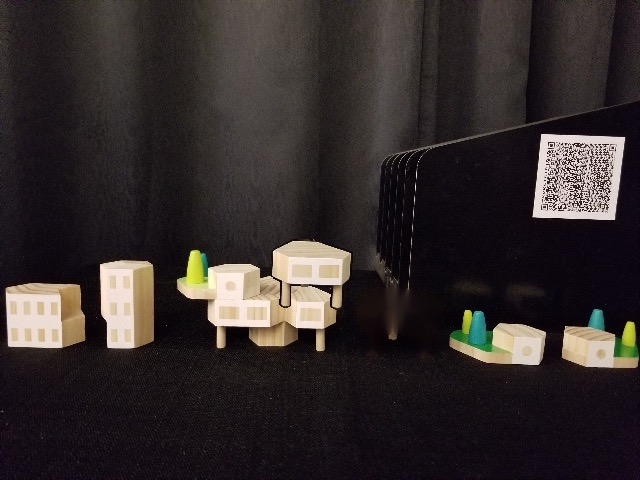}
\includegraphics[width=0.44\linewidth]{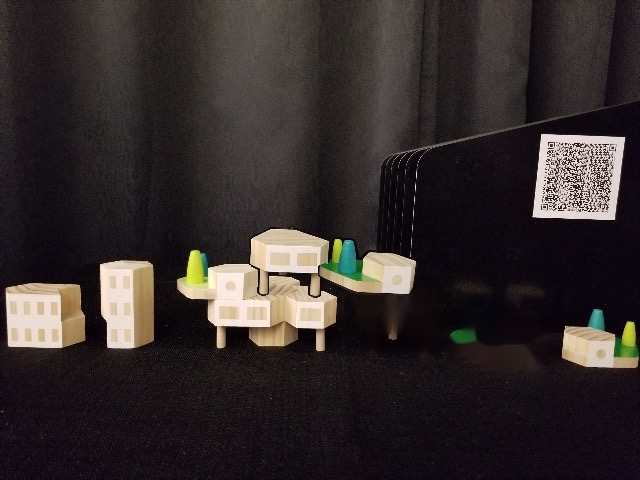}
\includegraphics[width=0.44\linewidth]{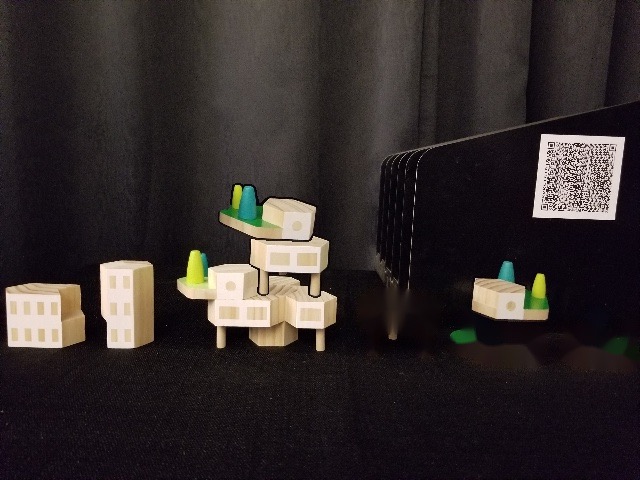}
\includegraphics[width=0.44\linewidth]{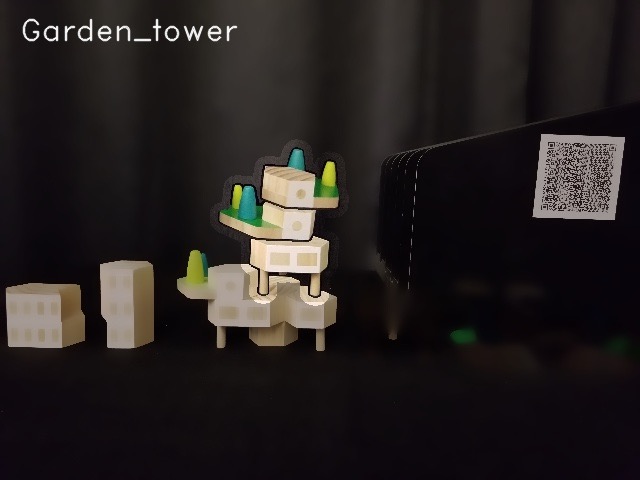}
\includegraphics[width=0.44\linewidth]{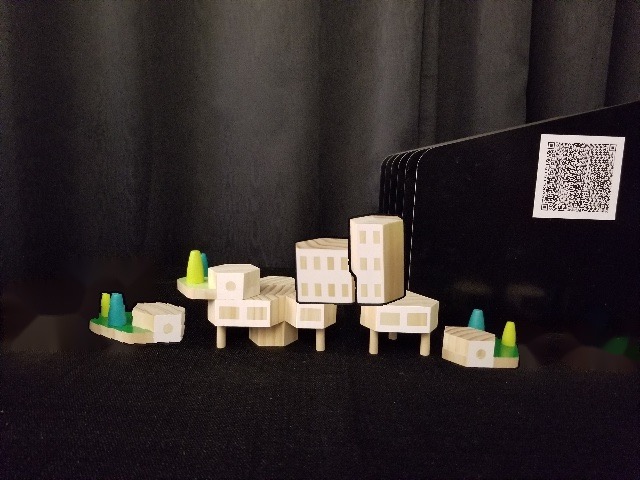}
\caption{A Blockitecture\textregistered\ set. The first five pictures show the process to build a garden tower, while the last picture shows the possibility of a different configuration. {\em Animation in supplemental video.}}
\label{fig:blocki}
\end{figure}

Building blocks are another example of design. 
Many designers and architects use blocks to develop ideas in 3D space. We demonstrate using a Blockitecture\textregistered\ set in Fig.~\ref{fig:blocki}. In Blockitecture, simply-shaped blocks 
can be rearranged in many different configurations. Using our system, the author can use a single scene to explore various configurations without physically rearranging the blocks. The author can simply take an image of the scene, and then in the authoring interface perform transformations on the blocks to demonstrate one design, and then restart the authoring process with the same image to move other blocks around to see a different design. Our system allows artists to explore physical designs without having to be physically alter the scene. This Blockitecture example can be expanded to larger scenarios where designers can take an image of a construction in progress, and then use our system to view various possibilities for the construction.

%% file: 6_evaluation.tex
\section{Evaluation}
\label{sec:6}

We evaluated our system with a user study. The goal of the study was to test whether people find our novel system easy to use and which aspects of the authoring interface are the most helpful in creating an animation. We also sought to explore how well people understand the animation decoded through the mobile app and how robust our system is to changes in environment lighting and camera position. People may be familiar with time-based media, authoring systems, and network-free communication as separate entities, but to our knowledge there is no existing system that embodies all three of the areas. Therefore, obtaining user feedback on authoring and consuming personalized time-based media in a network-free environment is important to determine whether such a system is effective for communication.

\subsection{User Study}
In a pilot study, we asked four users (two PhD students in computer science, an illustrator, and a student from a non-computing background) to test the authoring interface by creating a sample animation. They offered feedback on improvements for the interface, such as removing redundant components, adding the undo functionality, and hints about what to do at each step of the authoring process. 

For the formal user study, we chose five scenes from the previous section (Blockitecture, coffee maker, ink grinding, navigation on a map, and the Rubin vase) and an additional simple test scene. These five scenes cover various areas of application for our system, which would allow for more comprehensive results. We created an authored QR code for each scene, so that all subjects can view the same animations using the mobile app. 

We recruited 20 subjects of varying age and background. Each subject was shown a short instructional video demonstrating the authoring and consuming process. Participants were able to familiarize themselves with authoring and consuming animations using the test scene. They first used the mobile app on an Android smartphone (Google Pixel 2 XL, released in October 2017) to decode an animation from the test scene by registering QR code landmarks. Then they used the authoring interface on a laptop (MacBook Pro 13-inch, Late 2013) to recreate the animation based on what they understood from the animation decoded on the mobile device. We answered questions by the subjects as they worked through the test scene, to make sure they understood how to correctly use the mobile app and authoring interface. 

After understanding the authoring and consuming process through the test scene, participants repeated the same procedure for the five scenes set up for this user study. The procedure, again, involves first viewing an animation using the mobile app and then using the authoring interface to recreate the animation. All participants viewed the same animation for each scene. During these five scenes, participants were not allowed to ask us questions and did not receive input from us. The order of scenes for each participant was different, to account for participants getting better at using the system over time. For each scene, we recorded the time taken for subjects to register the view on the smartphone, as well as the time taken to recreate the animation using the authoring interface. After finishing all the scenes, subjects were asked to answer questions regarding how difficult the registration process was and how well they understood what was being conveyed in the animation (as a consumer), and how difficult they found it to create the animations and how satisfied they were with different components in the interface (as an author). We also collected background information about each subject's level of expertise in video editing, computer science, and computer graphics.


\subsection{Observations}
The following key observations are gained through the user study: 
\begin{itemize}
\item Experts and non-experts alike find the authoring interface easy to use, taking on average around 3 minutes per animation.
\item People are most satisfied with the interface components 2D transformation, color transformation, and annotation.
\item Registering QR landmarks using the smartphone is an unfamiliar process to many people at first, but they quickly master it and registration time decreases significantly.
\item Although registering QR landmarks can require a bit of effort, once correctly registered the decoded animation is easily understood by most viewers.
\item Decoded animations are understood by viewers throughout various times of the day with varying lighting conditions and slight variation in camera position.
\end{itemize}
	
First, our high-level analysis focuses on scores provided by the subjects. All the scores in the survey are on a discrete scale of 1 to 5. For rating the difficulty of registering QR codes, a rating of 5 was most difficult. For understanding decoded animations, 5 was easiest to understand. For evaluating interface components, 5 indicated the highest level of satisfaction. We explicitly explained rating semantics in the survey to avoid confusion.

The average time that people took to register a view was 41.79 seconds. People found it moderately difficult to register QR code landmarks (average rated difficulty: 3.45). Most people found it easy to understand what is being conveyed in the decoded animations (average rated level of understanding: 4.00). The average time taken by people to create an animation was 189.82 seconds. Most people found it easy to create an animation using the authoring interface (average rated difficulty: 2.19). In general, people were satisfied with the way they interacted with different interface components (segmentation tuning: 3.35, 2D transformation: 4.55, 3D transformation: 3.55, color transformation: 4.45, annotation: 4.65, animation preview: 4.35, undo: 4.00).
\begin{figure}[h]
\includegraphics[width=0.8\linewidth]{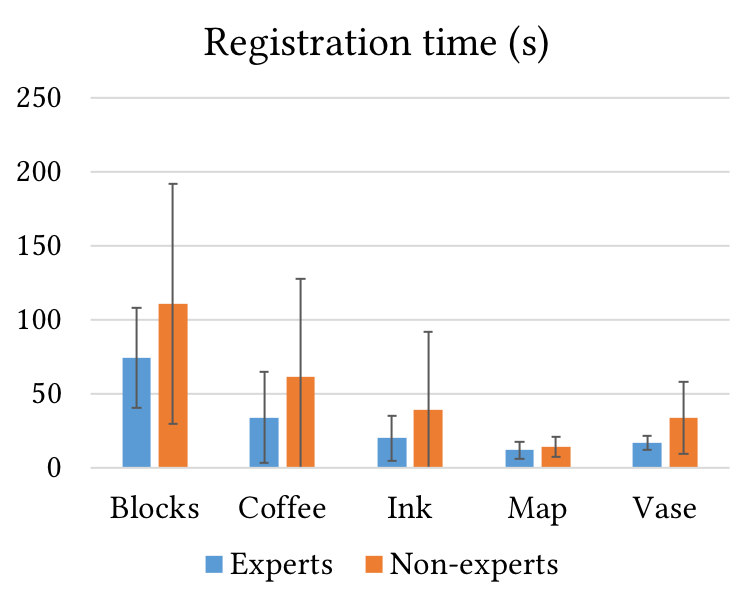}
\includegraphics[width=0.8\linewidth]{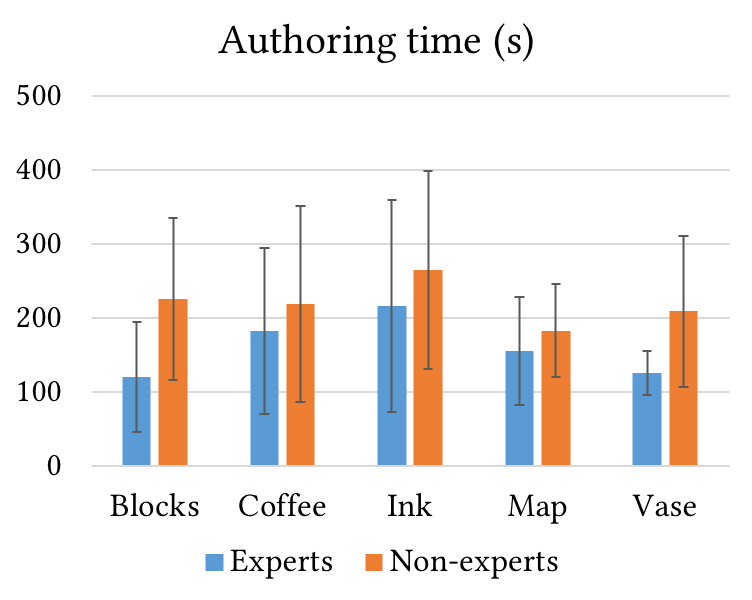}
\caption{Top: average time participants took to register a view. Bottom: average time participants took to create an animation.}
\label{fig:userstudy}
\end{figure}

We also wanted to explore whether there is any difference in behavior between expert and non-expert users. Based on the level of expertise in video editing, computer science, and computer graphics on a discrete scale of 1 to 5 with 5 being most experienced, we consider a participant an expert user if the three numbers add up to 8 or more, or a non-expert user otherwise. We had 10 expert users and 10 non-expert users. We recorded the registration time and authoring time for all users (Figure.~\ref{fig:userstudy}). Expert users took 20.51 seconds less on average to register a view, because they might have a better idea on how to move and tilt the smartphone in order to better register the QR code landmarks. The non-expert group had greater standard deviation, which suggests that their understanding of how to register a view varied among individuals. Expert users also took 59.97 seconds less on average to create an animation. However, if we compute the average animation creation time per individual, one-way analysis of variance (ANOVA) yields a $p$-value of 0.0748, greater than the common significance level of 0.05. This means no statistical difference was observed between experts and non-experts in terms of the authoring time. Given the fact that the average authoring time was about 3 minutes and all participants completed each animation in under 10 minutes, we conclude that non-experts are able to use the authoring interface almost as conveniently as expert users. 

We also asked participants for general comments on the AniCode framework regarding ease of use. The registration process was difficult for many, mainly because most of them had never done something similar. Registering a 3D scene took significantly longer than registering a 2D poster. However, we also observed that individuals improved rapidly. Once a subject knew how moving and tilting the smartphone affected the registration of QR code landmarks, a view could be registered in 10 seconds. For the authoring process, most participants liked the undo functionality and all the animation schemes except 3D transformation. Some of them had a hard time tuning the segmentation parameters or warping an object. This could be due to the fact that non-experts are not familiar with ideas of segmentation and homography. Participants also gave feedback on potential improvements for the interface, such as adding a scaling functionality, preserving segment selections between keyframes, and more flexibility in the list of thumbnails. Finally, participants mentioned interesting applications of the AniCode framework, including indoor navigation, interactive advertisement, museum displays, home decoration, AR treasure hunt, employee training, educational illustrations, etc. Overall, the participants enjoyed using the system both as a consumer and as an author. They found the system to be innovative and useful. User study resources such as the instructional video, survey, and raw data can be found in our supplementary materials.

From the sample applications and user study, we demonstrate that our system is able to handle various scenes from posters on a wall to objects displayed in the physical world. It gives reasonable results for scenes that consist of only pure color parts or specular and textured regions. To the best of our knowledge, AniCode is the first framework to communicate videos in a network-free environment, making the communication both private and personalized. The AniCode framework is innovative in four respects -- authors do not need to employ computer specialists to create expository media, the media incorporate new modes of presentation and interaction, internet communication is not required, and consumers are able to personalize their use of the media. The examples in the paper and supplementary portfolio show that our system is also versatile and applicable to a variety of areas such as education, consumer product documentation, cultural heritage preservation, creative art, and others that require explanations of objects and processes.

%% file: 7_limitations_and_future_work.tex
\section{Limitations and Future Work}
\label{sec:7}
\begin{figure}[b]
\includegraphics[width=0.49\linewidth]{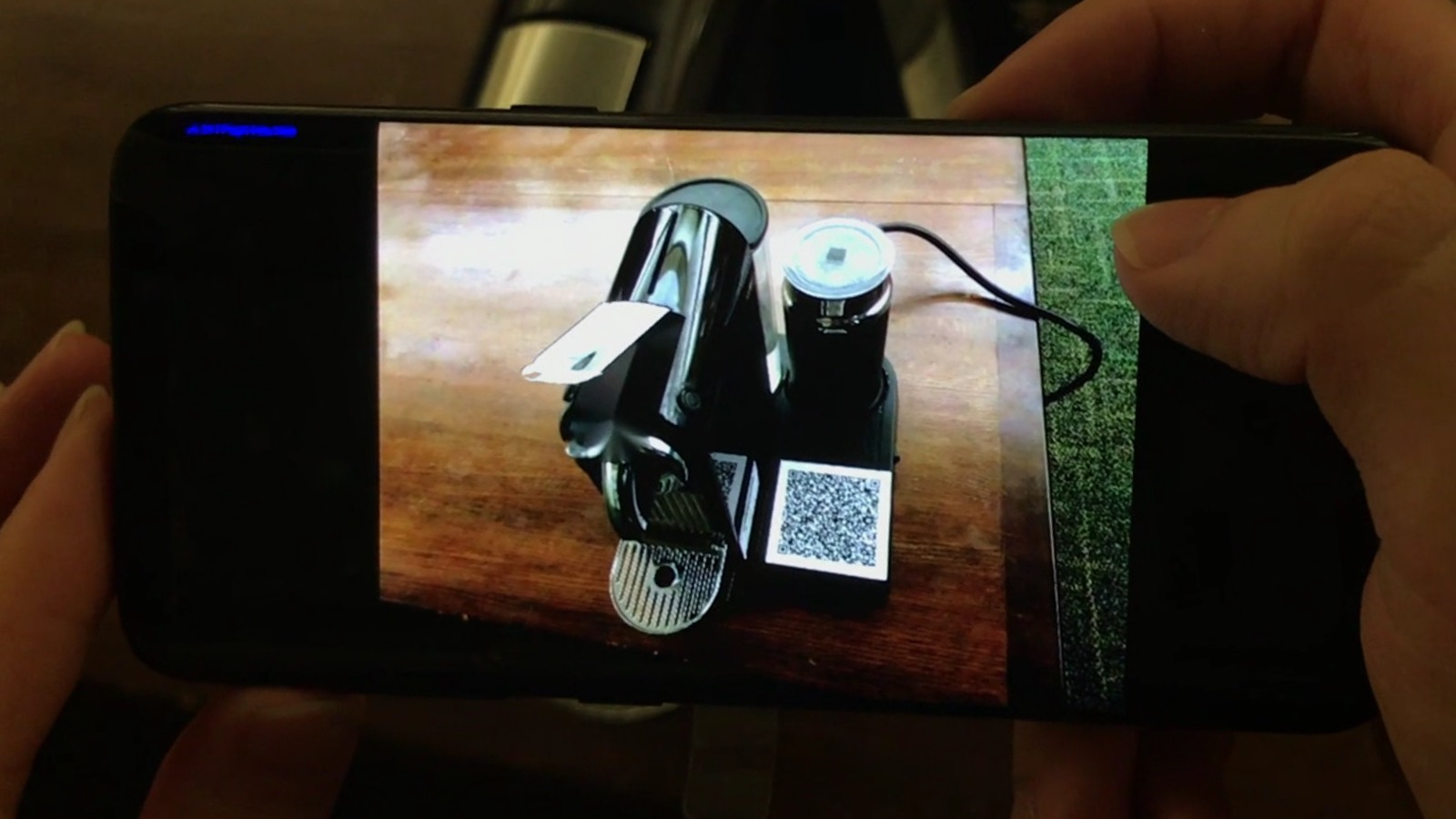}
\includegraphics[width=0.49\linewidth]{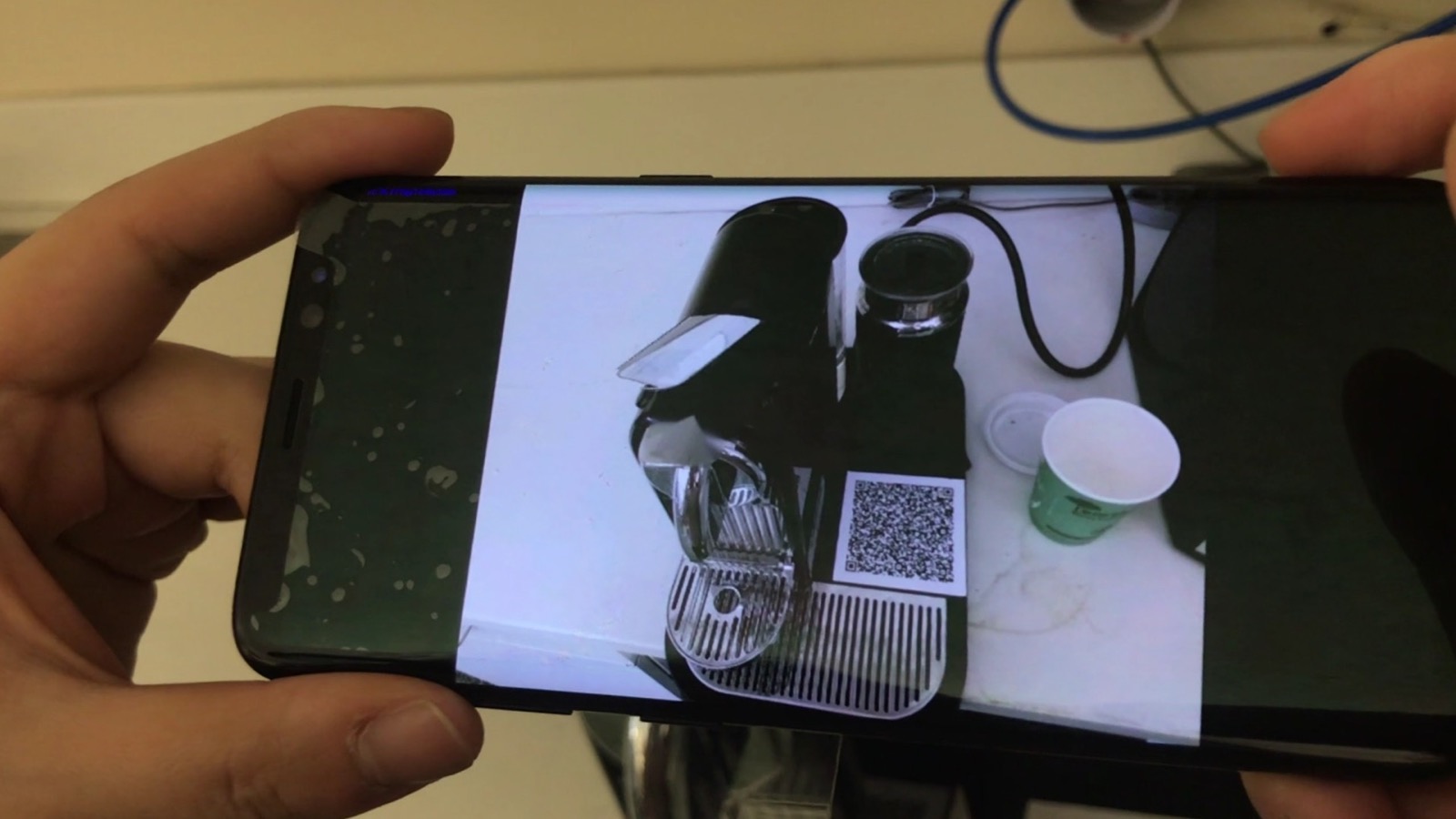}
\includegraphics[width=0.49\linewidth]{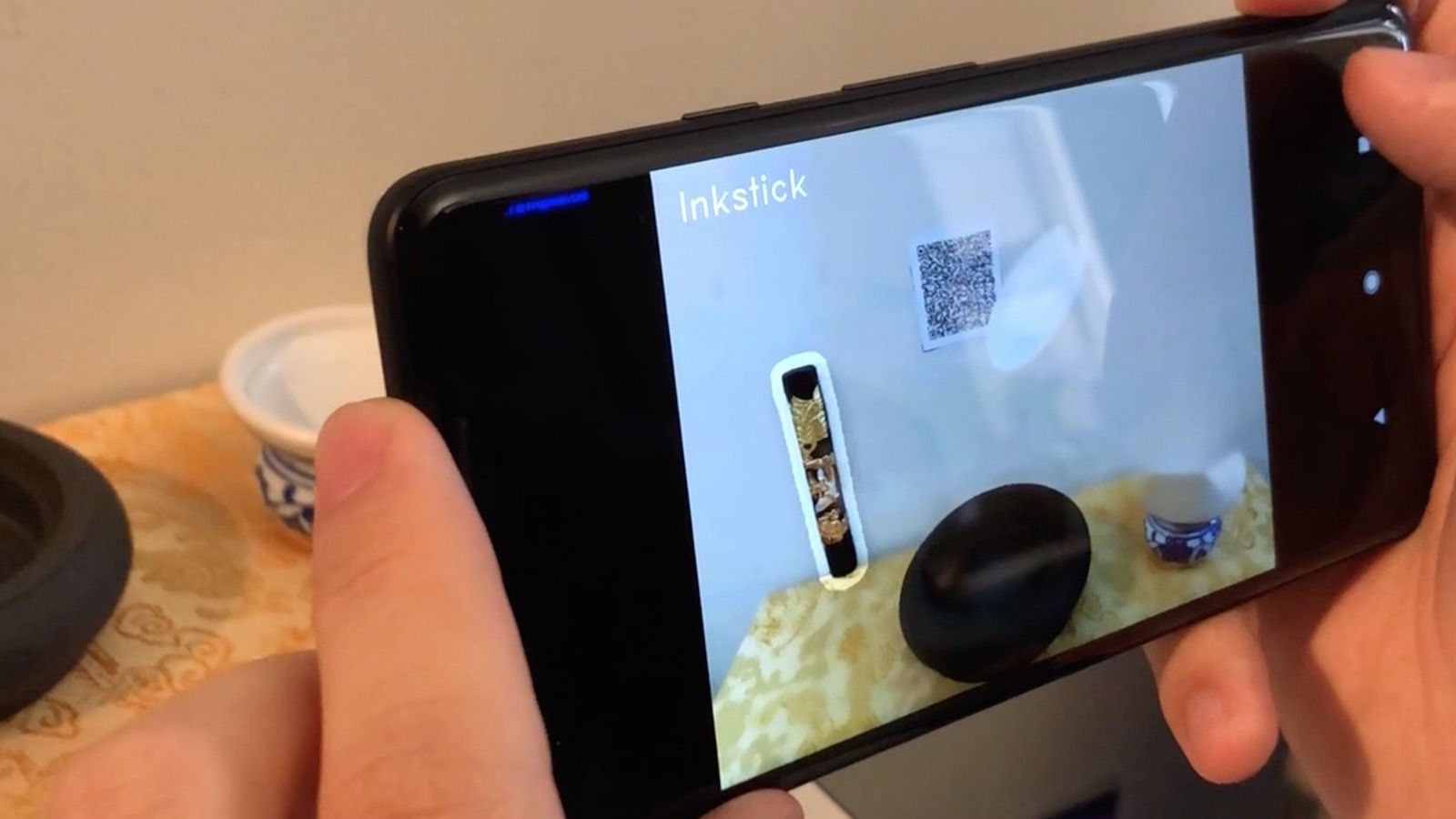}
\includegraphics[width=0.49\linewidth]{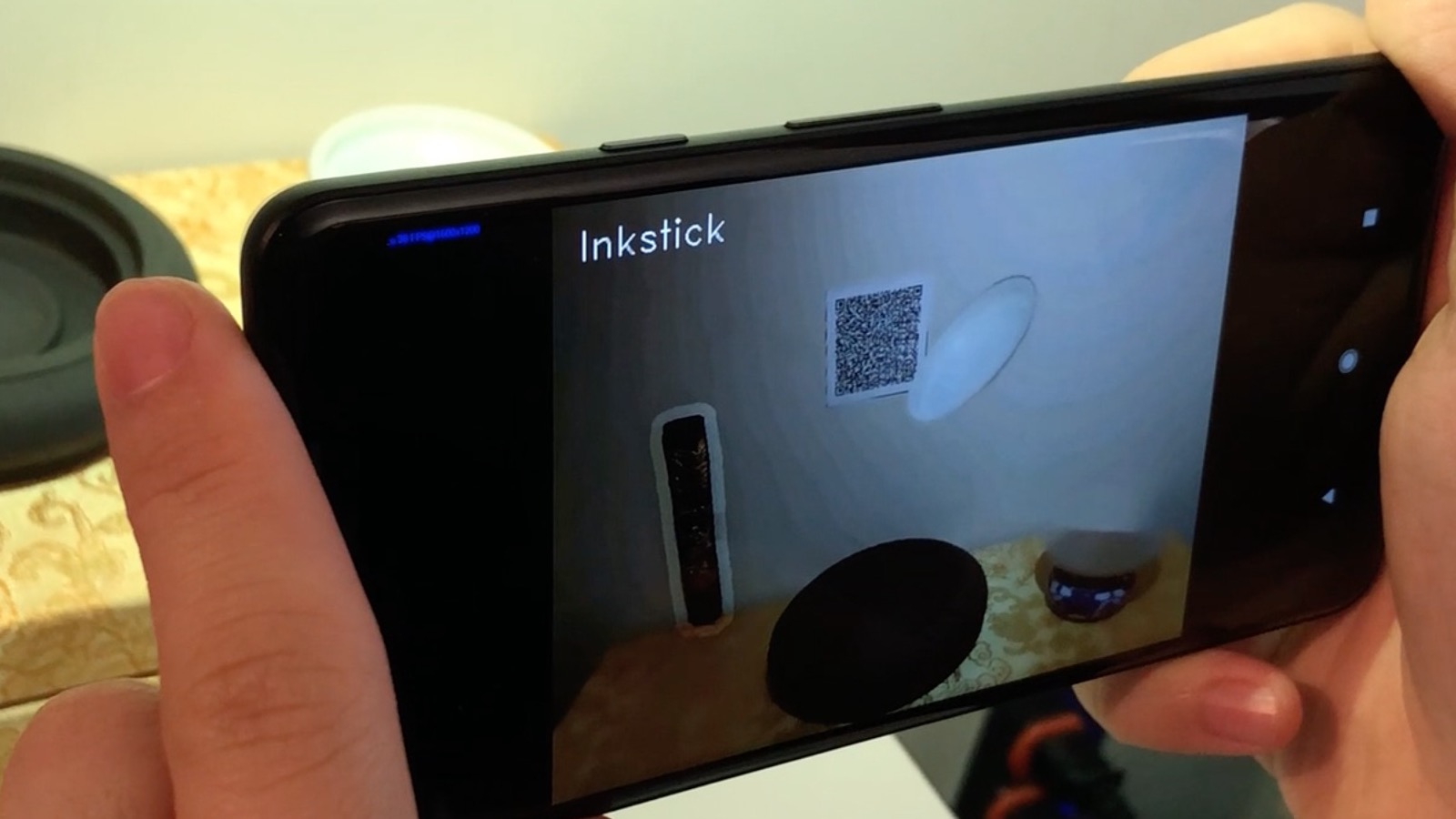}
\caption{Top: robustness to changes of the vantage point and environment. Bottom: robustness to different lighting conditions.}
\label{fig:robustness}
\end{figure}

The main limitation of our system is that the segmentation and inpainting results are not always perfect when there are extreme changes in the environment, because we choose the LSC and Telea algorithms as a tradeoff between performance and runtime on a mobile device. The consumer's view may not be exactly the same as the author's, and the object may be moved to another environment with different background and lighting. Figure.~\ref{fig:robustness} shows that our system is robust to reasonable changes of the vantage point and lighting conditions. The registration threshold we set in the mobile app is a tradeoff between registration difficulty and the quality of decoded animations. This means if the consumer takes a picture from a slightly different view, the algorithm usually succeeds in matching the ROIs and recovering a personalized animation. However, drastic changes of the view or lighting conditions can cause failure in ROI matching, which makes decoded animations sometimes confusing. Interestingly, we observed in the user study that participants were able to create the intended animation in the authoring interface even if some ROIs were mismatched. This shows that the author's ideas can still be effectively communicated using our system even when the visual effects are not perfect. The AniCode framework can be further improved by gaining design-related insights from more experiments such as determining the range of consumer camera angles allowed to decode a coherent animation and exploring easier ways for consumers to register a scene.

By design, AniCode only handles scenes whose parts are in a rigid configuration. If a consumer's scene has multiple objects that are in a configuration different from that of the author's, the correct animation cannot be decoded. In the future, this may be solved using multiple codes to support localization of objects in the scene. Furthermore, no additional objects can be introduced to the scene due to the limited capacity of QR codes and dependence on the consumer's picture to provide personalization. An improvement would be adding another animation scheme that enables introducing geometric primitives. Other ideas to enrich the animation schemes include rendering a contour around the highlighted ROI for annotation, adding image layer representations, and linking other pictures or videos that consumers can take. In particular, the AniCode framework can be generalized to use 3D information with recent smartphones that support capturing depth images. The current choice of QR code occupies a noticeable area in the scene. More alternatives such as aesthetically pleasing codes \cite{QRappearance}, watermarking techniques \cite{xiao2018fontcode}, and 3D fabricated unobtrusive tags \cite{li2017aircode} can be applied in the future. Other possible future work includes adopting image processing algorithms based on deep learning and developing new applications.


%% file: AniCode.bbl

\begin{thebibliography}{37}


\ifx \showCODEN    \undefined \def \showCODEN     #1{\unskip}     \fi
\ifx \showDOI      \undefined \def \showDOI       #1{#1}\fi
\ifx \showISBNx    \undefined \def \showISBNx     #1{\unskip}     \fi
\ifx \showISBNxiii \undefined \def \showISBNxiii  #1{\unskip}     \fi
\ifx \showISSN     \undefined \def \showISSN      #1{\unskip}     \fi
\ifx \showLCCN     \undefined \def \showLCCN      #1{\unskip}     \fi
\ifx \shownote     \undefined \def \shownote      #1{#1}          \fi
\ifx \showarticletitle \undefined \def \showarticletitle #1{#1}   \fi
\ifx \showURL      \undefined \def \showURL       {\relax}        \fi
\providecommand\bibfield[2]{#2}
\providecommand\bibinfo[2]{#2}
\providecommand\natexlab[1]{#1}
\providecommand\showeprint[2][]{arXiv:#2}

\bibitem[\protect\citeauthoryear{Achanta, Shaji, Smith, Lucchi, Fua, and
  S{\"u}sstrunk}{Achanta et~al\mbox{.}}{2012}]%
        {achanta2012slic}
\bibfield{author}{\bibinfo{person}{Radhakrishna Achanta}, \bibinfo{person}{Appu
  Shaji}, \bibinfo{person}{Kevin Smith}, \bibinfo{person}{Aurelien Lucchi},
  \bibinfo{person}{Pascal Fua}, {and} \bibinfo{person}{Sabine S{\"u}sstrunk}.}
  \bibinfo{year}{2012}\natexlab{}.
\newblock \showarticletitle{SLIC Superpixels Compared to State-of-the-Art
  Superpixel Methods}.
\newblock \bibinfo{journal}{{\em IEEE Transactions on Pattern Analysis and
  Machine Intelligence\/}} \bibinfo{volume}{34}, \bibinfo{number}{11}
  (\bibinfo{year}{2012}), \bibinfo{pages}{2274--2282}.
\newblock


\bibitem[\protect\citeauthoryear{Agrawala, Li, and Berthouzoz}{Agrawala
  et~al\mbox{.}}{2011}]%
        {Agrawala:2011:DPV:1924421.1924439}
\bibfield{author}{\bibinfo{person}{Maneesh Agrawala}, \bibinfo{person}{Wilmot
  Li}, {and} \bibinfo{person}{Floraine Berthouzoz}.}
  \bibinfo{year}{2011}\natexlab{}.
\newblock \showarticletitle{Design Principles for Visual Communication}.
\newblock \bibinfo{journal}{{\it Commun. ACM}} \bibinfo{volume}{54},
  \bibinfo{number}{4} (\bibinfo{date}{April} \bibinfo{year}{2011}),
  \bibinfo{pages}{60--69}.
\newblock
\showISSN{0001-0782}
\showDOI{%
\url{https://doi.org/10.1145/1924421.1924439}}


\bibitem[\protect\citeauthoryear{Agrawala, Phan, Heiser, Haymaker, Klingner,
  Hanrahan, and Tversky}{Agrawala et~al\mbox{.}}{2003}]%
        {Agrawala:2003:DES:882262.882352}
\bibfield{author}{\bibinfo{person}{Maneesh Agrawala}, \bibinfo{person}{Doantam
  Phan}, \bibinfo{person}{Julie Heiser}, \bibinfo{person}{John Haymaker},
  \bibinfo{person}{Jeff Klingner}, \bibinfo{person}{Pat Hanrahan}, {and}
  \bibinfo{person}{Barbara Tversky}.} \bibinfo{year}{2003}\natexlab{}.
\newblock \showarticletitle{Designing Effective Step-by-step Assembly
  Instructions}.
\newblock \bibinfo{journal}{{\em ACM Transactions on Graphics\/}}
  \bibinfo{volume}{22}, \bibinfo{number}{3} (\bibinfo{date}{July}
  \bibinfo{year}{2003}), \bibinfo{pages}{828--837}.
\newblock
\showISSN{0730-0301}
\showDOI{%
\url{https://doi.org/10.1145/882262.882352}}


\bibitem[\protect\citeauthoryear{Andolina, Pirrone, Russo, Sorce, and
  Gentile}{Andolina et~al\mbox{.}}{2012}]%
        {andolina2012exploitation}
\bibfield{author}{\bibinfo{person}{Salvatore Andolina}, \bibinfo{person}{Dario
  Pirrone}, \bibinfo{person}{Giuseppe Russo}, \bibinfo{person}{Salvatore
  Sorce}, {and} \bibinfo{person}{Antonio Gentile}.}
  \bibinfo{year}{2012}\natexlab{}.
\newblock \showarticletitle{{Exploitation of Mobile Access to Context-Based
  Information in Cultural Heritage Fruition}}. In \bibinfo{booktitle}{{\em
  Proceedings of the International Conference on Broadband, Wireless Computing,
  Communication and Applications}}. IEEE, \bibinfo{pages}{322--328}.
\newblock


\bibitem[\protect\citeauthoryear{Appiah}{Appiah}{2006}]%
        {appiah2006rich}
\bibfield{author}{\bibinfo{person}{Osei Appiah}.}
  \bibinfo{year}{2006}\natexlab{}.
\newblock \showarticletitle{Rich Media, Poor Media: The Impact of Audio/Video
  vs. Text/Picture Testimonial Ads on Browsers' Evaluations of Commercial Web
  Sites and Online Products}.
\newblock \bibinfo{journal}{{\em Journal of Current Issues \& Research in
  Advertising\/}} \bibinfo{volume}{28}, \bibinfo{number}{1}
  (\bibinfo{year}{2006}), \bibinfo{pages}{73--86}.
\newblock


\bibitem[\protect\citeauthoryear{Ashok}{Ashok}{2014}]%
        {ashok2014}
\bibfield{author}{\bibinfo{person}{Ashwin Ashok}.}
  \bibinfo{year}{2014}\natexlab{}.
\newblock {\em \bibinfo{title}{Design, Modeling, and Analysis of Visual MIMO
  Communication}}.
\newblock \bibinfo{thesistype}{Ph.D. Dissertation}. \bibinfo{school}{Rutgers
  The State University of New Jersey-New Brunswick}.
\newblock


\bibitem[\protect\citeauthoryear{Barak, Ashkar, and Dori}{Barak
  et~al\mbox{.}}{2011}]%
        {barak2011learning}
\bibfield{author}{\bibinfo{person}{Miri Barak}, \bibinfo{person}{Tamar Ashkar},
  {and} \bibinfo{person}{Yehudit~J. Dori}.} \bibinfo{year}{2011}\natexlab{}.
\newblock \showarticletitle{Teaching Science via Animated Movies: Its Effect on
  Students' Thinking and Motivation}.
\newblock \bibinfo{journal}{{\em Computers \& Education\/}}
  \bibinfo{volume}{56}, \bibinfo{number}{3} (\bibinfo{year}{2011}),
  \bibinfo{pages}{839--846}.
\newblock


\bibitem[\protect\citeauthoryear{Carter, Cooper, Adcock, and Branham}{Carter
  et~al\mbox{.}}{2014}]%
        {Carter2014}
\bibfield{author}{\bibinfo{person}{Scott Carter}, \bibinfo{person}{Matthew
  Cooper}, \bibinfo{person}{John Adcock}, {and} \bibinfo{person}{Stacy
  Branham}.} \bibinfo{year}{2014}\natexlab{}.
\newblock \showarticletitle{Tools to Support Expository Video Capture and
  Access}.
\newblock \bibinfo{journal}{{\em Education and Information Technologies\/}}
  \bibinfo{volume}{19}, \bibinfo{number}{3} (\bibinfo{date}{Sep}
  \bibinfo{year}{2014}), \bibinfo{pages}{637--654}.
\newblock


\bibitem[\protect\citeauthoryear{Carter, Qvarfordt, Cooper, and
  M{\"a}kel{\"a}}{Carter et~al\mbox{.}}{2015}]%
        {carter2015creating}
\bibfield{author}{\bibinfo{person}{Scott Carter}, \bibinfo{person}{Pernilla
  Qvarfordt}, \bibinfo{person}{Matthew Cooper}, {and} \bibinfo{person}{Ville
  M{\"a}kel{\"a}}.} \bibinfo{year}{2015}\natexlab{}.
\newblock \showarticletitle{Creating Tutorials with Web-Based Authoring and
  Heads-Up Capture}.
\newblock \bibinfo{journal}{{\em IEEE Pervasive Computing\/}}
  \bibinfo{volume}{14}, \bibinfo{number}{3} (\bibinfo{year}{2015}),
  \bibinfo{pages}{44--52}.
\newblock


\bibitem[\protect\citeauthoryear{Chang, Chu, and Mitra}{Chang
  et~al\mbox{.}}{2016}]%
        {chang2016interactive}
\bibfield{author}{\bibinfo{person}{Chia-Sheng Chang}, \bibinfo{person}{Hung-Kuo
  Chu}, {and} \bibinfo{person}{Niloy~J. Mitra}.}
  \bibinfo{year}{2016}\natexlab{}.
\newblock \showarticletitle{Interactive Videos: Plausible Video Editing using
  Sparse Structure Points}.
\newblock \bibinfo{journal}{{\em Computer Graphics Forum\/}}
  \bibinfo{volume}{35}, \bibinfo{number}{2} (\bibinfo{year}{2016}),
  \bibinfo{pages}{489--500}.
\newblock
\showDOI{%
\url{https://doi.org/10.1111/cgf.12849}}


\bibitem[\protect\citeauthoryear{Cho, Wu, Xu, and Zhang}{Cho
  et~al\mbox{.}}{2016}]%
        {cho2016content}
\bibfield{author}{\bibinfo{person}{Nan-Hung Cho}, \bibinfo{person}{Qiang Wu},
  \bibinfo{person}{Jingsong Xu}, {and} \bibinfo{person}{Jian Zhang}.}
  \bibinfo{year}{2016}\natexlab{}.
\newblock \showarticletitle{Content Authoring Using Single Image in Urban
  Environments for Augmented Reality}. In \bibinfo{booktitle}{{\em Proceedings
  of the International Conference on Digital Image Computing: Techniques and
  Applications}}. IEEE, \bibinfo{pages}{1--7}.
\newblock


\bibitem[\protect\citeauthoryear{Chu, Bryan, Shih, Ferrer, and Ma}{Chu
  et~al\mbox{.}}{2017}]%
        {chu2017navigable}
\bibfield{author}{\bibinfo{person}{Jacqueline Chu}, \bibinfo{person}{Chris
  Bryan}, \bibinfo{person}{Min Shih}, \bibinfo{person}{Leonardo Ferrer}, {and}
  \bibinfo{person}{Kwan-Liu Ma}.} \bibinfo{year}{2017}\natexlab{}.
\newblock \showarticletitle{Navigable Videos for Presenting Scientific Data on
  Affordable Head-Mounted Displays}. In \bibinfo{booktitle}{{\em Proceedings of
  the ACM Conference on Multimedia Systems}}. ACM, \bibinfo{pages}{250--260}.
\newblock


\bibitem[\protect\citeauthoryear{Clarine}{Clarine}{2016}]%
        {clarine}
\bibfield{author}{\bibinfo{person}{Brenna Clarine}.}
  \bibinfo{year}{2016}\natexlab{}.
\newblock \bibinfo{title}{11 Reasons Why Video is Better Than Any Other
  Medium}.
\newblock
  \bibinfo{howpublished}{\url{http://www.advancedwebranking.com/blog/11-reasons-why-video-is-better}}.
    (\bibinfo{year}{2016}).
\newblock


\bibitem[\protect\citeauthoryear{Feiner and McKeown}{Feiner and
  McKeown}{1991}]%
        {feiner1991automating}
\bibfield{author}{\bibinfo{person}{Steven~K. Feiner} {and}
  \bibinfo{person}{Kathleen~R. McKeown}.} \bibinfo{year}{1991}\natexlab{}.
\newblock \showarticletitle{Automating the Generation of Coordinated Multimedia
  Explanations}.
\newblock \bibinfo{journal}{{\em Computer\/}} \bibinfo{volume}{24},
  \bibinfo{number}{10} (\bibinfo{year}{1991}), \bibinfo{pages}{33--41}.
\newblock


\bibitem[\protect\citeauthoryear{Fidas, Sintoris, Yiannoutsou, and
  Avouris}{Fidas et~al\mbox{.}}{2015}]%
        {7388029}
\bibfield{author}{\bibinfo{person}{Christos Fidas}, \bibinfo{person}{Christos
  Sintoris}, \bibinfo{person}{Nikoleta Yiannoutsou}, {and}
  \bibinfo{person}{Nikolaos Avouris}.} \bibinfo{year}{2015}\natexlab{}.
\newblock \showarticletitle{A Survey on Tools for End User Authoring of Mobile
  Applications for Cultural Heritage}. In \bibinfo{booktitle}{{\em Proceedings
  of the International Conference on Information, Intelligence, Systems and
  Applications}}. \bibinfo{pages}{1--5}.
\newblock
\showDOI{%
\url{https://doi.org/10.1109/IISA.2015.7388029}}


\bibitem[\protect\citeauthoryear{Hern}{Hern}{2018}]%
        {Strava}
\bibfield{author}{\bibinfo{person}{Alex Hern}.}
  \bibinfo{year}{2018}\natexlab{}.
\newblock \showarticletitle{Fitness Tracking App Strava Gives Away Location of
  Secret US Army Bases}.
\newblock \bibinfo{journal}{{\em The Guardian\/}} (\bibinfo{year}{2018}).
\newblock
\showURL{%
\url{https://www.theguardian.com/world/2018/jan/28/fitness-tracking-app-gives-away-location-of-secret-us-army-bases}}


\bibitem[\protect\citeauthoryear{Kaplan and Haenlein}{Kaplan and
  Haenlein}{2010}]%
        {kaplan2010users}
\bibfield{author}{\bibinfo{person}{Andreas~M Kaplan} {and}
  \bibinfo{person}{Michael Haenlein}.} \bibinfo{year}{2010}\natexlab{}.
\newblock \showarticletitle{Users of the World, Unite! The Challenges and
  Opportunities of Social Media}.
\newblock \bibinfo{journal}{{\em Business horizons\/}} \bibinfo{volume}{53},
  \bibinfo{number}{1} (\bibinfo{year}{2010}), \bibinfo{pages}{59--68}.
\newblock


\bibitem[\protect\citeauthoryear{Karat, Pinhanez, Karat, Arora, and
  Vergo}{Karat et~al\mbox{.}}{2001}]%
        {karat2001less}
\bibfield{author}{\bibinfo{person}{Clare-Marie Karat}, \bibinfo{person}{Claudio
  Pinhanez}, \bibinfo{person}{John Karat}, \bibinfo{person}{Renee Arora}, {and}
  \bibinfo{person}{John Vergo}.} \bibinfo{year}{2001}\natexlab{}.
\newblock \showarticletitle{Less Clicking, More Watching: Results of the
  Iterative Design and Evaluation of Entertaining Web Experiences}. In
  \bibinfo{booktitle}{{\em Proceedings of the IFIP TC13 International
  Conference on Human-Computer Interaction}}. \bibinfo{pages}{455--463}.
\newblock


\bibitem[\protect\citeauthoryear{Li, Nair, Nayar, and Zheng}{Li
  et~al\mbox{.}}{2017}]%
        {li2017aircode}
\bibfield{author}{\bibinfo{person}{Dingzeyu Li}, \bibinfo{person}{Avinash~S.
  Nair}, \bibinfo{person}{Shree~K. Nayar}, {and} \bibinfo{person}{Changxi
  Zheng}.} \bibinfo{year}{2017}\natexlab{}.
\newblock \showarticletitle{AirCode: Unobtrusive Physical Tags for Digital
  Fabrication}. In \bibinfo{booktitle}{{\em Proceedings of the ACM Symposium on
  User Interface Software and Technology}}. ACM, \bibinfo{pages}{449--460}.
\newblock


\bibitem[\protect\citeauthoryear{Li and Chen}{Li and Chen}{2015}]%
        {li2015superpixel}
\bibfield{author}{\bibinfo{person}{Zhengqin Li} {and}
  \bibinfo{person}{Jiansheng Chen}.} \bibinfo{year}{2015}\natexlab{}.
\newblock \showarticletitle{Superpixel Segmentation Using Linear Spectral
  Clustering}. In \bibinfo{booktitle}{{\em Proceedings of the IEEE Conference
  on Computer Vision and Pattern Recognition}}. \bibinfo{pages}{1356--1363}.
\newblock


\bibitem[\protect\citeauthoryear{Liao, Hsu, and Ma}{Liao et~al\mbox{.}}{2014}]%
        {Liao2014}
\bibfield{author}{\bibinfo{person}{Isaac Liao}, \bibinfo{person}{Wei-Hsien
  Hsu}, {and} \bibinfo{person}{Kwan-Liu Ma}.} \bibinfo{year}{2014}\natexlab{}.
\newblock \showarticletitle{Storytelling via Navigation: A Novel Approach to
  Animation for Scientific Visualization}. In \bibinfo{booktitle}{{\em
  Proceedings of the International Symposium on Smart Graphics}},
  \bibfield{editor}{\bibinfo{person}{Marc Christie} {and}
  \bibinfo{person}{Tsai-Yen Li}} (Eds.). \bibinfo{publisher}{Springer
  International Publishing}, \bibinfo{address}{Cham}, \bibinfo{pages}{1--14}.
\newblock


\bibitem[\protect\citeauthoryear{Lin, Hu, Lee, and Lee}{Lin
  et~al\mbox{.}}{2015}]%
        {QRappearance}
\bibfield{author}{\bibinfo{person}{Shih-Syun Lin}, \bibinfo{person}{Min-Chun
  Hu}, \bibinfo{person}{Chien-Han Lee}, {and} \bibinfo{person}{Tong-Yee Lee}.}
  \bibinfo{year}{2015}\natexlab{}.
\newblock \showarticletitle{Efficient QR Code Beautification With High Quality
  Visual Content}.
\newblock \bibinfo{journal}{{\em IEEE Transactions on Multimedia\/}}
  \bibinfo{volume}{17}, \bibinfo{number}{9} (\bibinfo{date}{Sept}
  \bibinfo{year}{2015}), \bibinfo{pages}{1515--1524}.
\newblock
\showISSN{1520-9210}
\showDOI{%
\url{https://doi.org/10.1109/TMM.2015.2437711}}


\bibitem[\protect\citeauthoryear{McKercher and Du~Cros}{McKercher and
  Du~Cros}{2002}]%
        {murcher}
\bibfield{author}{\bibinfo{person}{Bob McKercher} {and} \bibinfo{person}{Hilary
  Du~Cros}.} \bibinfo{year}{2002}\natexlab{}.
\newblock \bibinfo{booktitle}{{\em Cultural Tourism: The Partnership Between
  Tourism and Cultural Heritage Management}}.
\newblock \bibinfo{publisher}{Routledge}.
\newblock


\bibitem[\protect\citeauthoryear{{OpenCV team}}{{OpenCV team}}{2017}]%
        {opencv}
\bibfield{author}{\bibinfo{person}{{OpenCV team}}.}
  \bibinfo{year}{2017}\natexlab{}.
\newblock \bibinfo{title}{{OpenCV for Android SDK}}.
\newblock \bibinfo{howpublished}{\url{https://opencv.org/platforms/android}}.
  (\bibinfo{year}{2017}).
\newblock


\bibitem[\protect\citeauthoryear{Owen, Switkin, and team}{Owen
  et~al\mbox{.}}{2017}]%
        {zxing}
\bibfield{author}{\bibinfo{person}{Sean Owen}, \bibinfo{person}{Daniel
  Switkin}, {and} \bibinfo{person}{ZXing team}.}
  \bibinfo{year}{2017}\natexlab{}.
\newblock \bibinfo{title}{{ZXing} Barcode Scanning Library}.
\newblock \bibinfo{howpublished}{\url{https://github.com/zxing/zxing}}.
  (\bibinfo{year}{2017}).
\newblock


\bibitem[\protect\citeauthoryear{Parent}{Parent}{2012}]%
        {parent2012computer}
\bibfield{author}{\bibinfo{person}{Rick Parent}.}
  \bibinfo{year}{2012}\natexlab{}.
\newblock \bibinfo{booktitle}{{\em Computer Animation: Algorithms and
  Techniques}}.
\newblock \bibinfo{publisher}{Newnes}.
\newblock


\bibitem[\protect\citeauthoryear{Revell}{Revell}{2017}]%
        {IOXP}
\bibfield{author}{\bibinfo{person}{Timothy Revell}.}
  \bibinfo{year}{2017}\natexlab{}.
\newblock \showarticletitle{App Creates Augmented-Reality Tutorials from Normal
  Videos}.
\newblock \bibinfo{journal}{{\em New Scientist\/}} (\bibinfo{year}{2017}).
\newblock
\showURL{%
\url{https://www.newscientist.com/article/2146850-app-creates-augmented-reality-tutorials-from-normal-videos}}


\bibitem[\protect\citeauthoryear{Schnotz}{Schnotz}{2002}]%
        {Schnotz2002}
\bibfield{author}{\bibinfo{person}{Wolfgang Schnotz}.}
  \bibinfo{year}{2002}\natexlab{}.
\newblock \showarticletitle{Commentary: Towards an Integrated View of Learning
  from Text and Visual Displays}.
\newblock \bibinfo{journal}{{\em Educational Psychology Review\/}}
  \bibinfo{volume}{14}, \bibinfo{number}{1} (\bibinfo{date}{01 Mar}
  \bibinfo{year}{2002}), \bibinfo{pages}{101--120}.
\newblock
\showISSN{1573-336X}
\showDOI{%
\url{https://doi.org/10.1023/A:1013136727916}}


\bibitem[\protect\citeauthoryear{Telea}{Telea}{2004}]%
        {telea2004image}
\bibfield{author}{\bibinfo{person}{Alexandru Telea}.}
  \bibinfo{year}{2004}\natexlab{}.
\newblock \showarticletitle{An Image Inpainting Technique Based on the Fast
  Marching Method}.
\newblock \bibinfo{journal}{{\em Journal of Graphics Tools\/}}
  \bibinfo{volume}{9}, \bibinfo{number}{1} (\bibinfo{year}{2004}),
  \bibinfo{pages}{23--34}.
\newblock


\bibitem[\protect\citeauthoryear{Upson, Faulhaber, Kamins, Laidlaw, Schlegel,
  Vroom, Gurwitz, and Van~Dam}{Upson et~al\mbox{.}}{1989}]%
        {31462}
\bibfield{author}{\bibinfo{person}{Craig Upson}, \bibinfo{person}{Thomas
  Faulhaber, Jr.}, \bibinfo{person}{David Kamins}, \bibinfo{person}{David
  Laidlaw}, \bibinfo{person}{David Schlegel}, \bibinfo{person}{Jefrey Vroom},
  \bibinfo{person}{Robert Gurwitz}, {and} \bibinfo{person}{Andries Van~Dam}.}
  \bibinfo{year}{1989}\natexlab{}.
\newblock \showarticletitle{The Application Visualization System: A
  Computational Environment for Scientific Visualization}.
\newblock \bibinfo{journal}{{\em IEEE Computer Graphics and Applications\/}}
  \bibinfo{volume}{9}, \bibinfo{number}{4} (\bibinfo{date}{July}
  \bibinfo{year}{1989}), \bibinfo{pages}{30--42}.
\newblock
\showISSN{0272-1716}
\showDOI{%
\url{https://doi.org/10.1109/38.31462}}


\bibitem[\protect\citeauthoryear{Van~den Bergh, Boix, Roig, de~Capitani, and
  Van~Gool}{Van~den Bergh et~al\mbox{.}}{2012}]%
        {van2012seeds}
\bibfield{author}{\bibinfo{person}{Michael Van~den Bergh},
  \bibinfo{person}{Xavier Boix}, \bibinfo{person}{Gemma Roig},
  \bibinfo{person}{Benjamin de Capitani}, {and} \bibinfo{person}{Luc
  Van~Gool}.} \bibinfo{year}{2012}\natexlab{}.
\newblock \showarticletitle{SEEDS: Superpixels Extracted via Energy-Driven
  Sampling}. In \bibinfo{booktitle}{{\em Proceedings of the European Conference
  on Computer Vision}}. Springer, \bibinfo{pages}{13--26}.
\newblock


\bibitem[\protect\citeauthoryear{Wouters, Paas, and van
  Merri{\"e}nboer}{Wouters et~al\mbox{.}}{2008}]%
        {wouters2008optimize}
\bibfield{author}{\bibinfo{person}{Pieter Wouters}, \bibinfo{person}{Fred
  Paas}, {and} \bibinfo{person}{Jeroen~J.G. van Merri{\"e}nboer}.}
  \bibinfo{year}{2008}\natexlab{}.
\newblock \showarticletitle{How to Optimize Learning from Animated Models: A
  Review of Guidelines Based on Cognitive Load}.
\newblock \bibinfo{journal}{{\em Review of Educational Research\/}}
  \bibinfo{volume}{78}, \bibinfo{number}{3} (\bibinfo{year}{2008}),
  \bibinfo{pages}{645--675}.
\newblock


\bibitem[\protect\citeauthoryear{Xiao, Zhang, and Zheng}{Xiao
  et~al\mbox{.}}{2018}]%
        {xiao2018fontcode}
\bibfield{author}{\bibinfo{person}{Chang Xiao}, \bibinfo{person}{Cheng Zhang},
  {and} \bibinfo{person}{Changxi Zheng}.} \bibinfo{year}{2018}\natexlab{}.
\newblock \showarticletitle{FontCode: Embedding Information in Text Documents
  using Glyph Perturbation}.
\newblock \bibinfo{journal}{{\em ACM Transactions on Graphics\/}}
  \bibinfo{volume}{37}, \bibinfo{number}{2} (\bibinfo{year}{2018}),
  \bibinfo{pages}{15}.
\newblock


\bibitem[\protect\citeauthoryear{Yeshurun and Carrasco}{Yeshurun and
  Carrasco}{1998}]%
        {yeshurun1998attention}
\bibfield{author}{\bibinfo{person}{Yaffa Yeshurun} {and}
  \bibinfo{person}{Marisa Carrasco}.} \bibinfo{year}{1998}\natexlab{}.
\newblock \showarticletitle{Attention Improves or Impairs Visual Performance by
  Enhancing Spatial Resolution}.
\newblock \bibinfo{journal}{{\em Nature\/}} \bibinfo{volume}{396},
  \bibinfo{number}{6706} (\bibinfo{year}{1998}), \bibinfo{pages}{72--75}.
\newblock


\bibitem[\protect\citeauthoryear{Yuan, Dana, Varga, Ashok, Gruteser, and
  Mandayam}{Yuan et~al\mbox{.}}{2011}]%
        {ashok2011}
\bibfield{author}{\bibinfo{person}{W. Yuan}, \bibinfo{person}{K. Dana},
  \bibinfo{person}{M. Varga}, \bibinfo{person}{A. Ashok}, \bibinfo{person}{M.
  Gruteser}, {and} \bibinfo{person}{N. Mandayam}.}
  \bibinfo{year}{2011}\natexlab{}.
\newblock \showarticletitle{Computer Vision Methods for Visual MIMO Optical
  System}. In \bibinfo{booktitle}{{\em CVPR Workshops}}.
  \bibinfo{pages}{37--43}.
\newblock
\showISSN{2160-7508}
\showDOI{%
\url{https://doi.org/10.1109/CVPRW.2011.5981735}}


\bibitem[\protect\citeauthoryear{Yue, Yang, Ren, and Wang}{Yue
  et~al\mbox{.}}{2017}]%
        {yue2017scenectrl}
\bibfield{author}{\bibinfo{person}{Ya-Ting Yue}, \bibinfo{person}{Yong-Liang
  Yang}, \bibinfo{person}{Gang Ren}, {and} \bibinfo{person}{Wenping Wang}.}
  \bibinfo{year}{2017}\natexlab{}.
\newblock \showarticletitle{SceneCtrl: Mixed Reality Enhancement via Efficient
  Scene Editing}. In \bibinfo{booktitle}{{\em Proceedings of the ACM Symposium
  on User Interface Software and Technology}}. ACM, \bibinfo{pages}{427--436}.
\newblock


\bibitem[\protect\citeauthoryear{Zheng, Chen, Cheng, Zhou, Hu, and Mitra}{Zheng
  et~al\mbox{.}}{2012}]%
        {zheng2012interactive}
\bibfield{author}{\bibinfo{person}{Youyi Zheng}, \bibinfo{person}{Xiang Chen},
  \bibinfo{person}{Ming-Ming Cheng}, \bibinfo{person}{Kun Zhou},
  \bibinfo{person}{Shi-Min Hu}, {and} \bibinfo{person}{Niloy~J. Mitra}.}
  \bibinfo{year}{2012}\natexlab{}.
\newblock \showarticletitle{Interactive Images: Cuboid Proxies for Smart Image
  Manipulation}.
\newblock \bibinfo{journal}{{\em ACM Transactions on Graphics\/}}
  \bibinfo{volume}{31}, \bibinfo{number}{4} (\bibinfo{year}{2012}),
  \bibinfo{pages}{99}.
\newblock


\end{thebibliography}
